\documentclass[fleqn,usenatbib]{mnras}

\usepackage{newtxtext,newtxmath}
\usepackage[T1]{fontenc}

\DeclareRobustCommand{\VAN}[3]{#2}
\let\VANthebibliography\thebibliography
\def\thebibliography{\DeclareRobustCommand{\VAN}[3]{##3}\VANthebibliography}

\usepackage{graphicx}	% Including figure files
\usepackage{amsmath}	% Advanced maths commands
\usepackage{orcidlink} % for linking orcid's
\title[Tidal features within groups and clusters]{Tidal features around simulated groups and cluster galaxies: Enhancement and suppression of merger events through environment in LSST-like mock observations}
\author[Khalid et al.]{
Aman Khalid\textsuperscript{\orcidlink{0000-0003-1302-4426}},$^{1,2}$\thanks{E-mail: aman.khalid@unsw.edu.au}
Sarah Brough\textsuperscript{\orcidlink{0000-0002-9796-1363}},$^{1,2}$
Garreth Martin\textsuperscript{\orcidlink{0000-0003-2939-8668}},$^{3}$
Lucas C. Kimmig\textsuperscript{\orcidlink{0009-0006-8337-8712}},$^{4}$ 
Rhea-Silvia Remus\textsuperscript{\orcidlink{0009-0008-9260-7278}},$^{4}$ 
\newauthor
Claudia del P. Lagos\textsuperscript{\orcidlink{0000-0003-3021-8564}},$^{2,5}$ 
Lucas M. Valenzuela\textsuperscript{\orcidlink{0000-0002-7972-9675}},$^{4}$
Ruby J. Wright\textsuperscript{\orcidlink{0000-0002-1367-0949 }}$^{5}$
\\
$^{1}$School of Physics, University of New South Wales, NSW 2052, Australia\\
$^{2}$ARC Centre of Excellence for All Sky Astrophysics in 3 Dimensions (ASTRO 3D)\\
$^{3}$School of Physics and Astronomy, University of Nottingham, University Park, Nottingham NG7 2RD, UK\\
$^{4}$Universitäts-Sternwarte München, Fakultät für Physik, LMU München, Scheinerstr. 1, D-81679 München, Germany\\
$^{5}$International Centre for Radio Astronomy Research, The University of Western Australia, 35 Stirling Highway, Crawley, WA 6009, Australia\\
\\
}

\date{Accepted XXX. Received YYY; in original form ZZZ}

\pubyear{\the\year{}}

\begin{document}
\renewcommand{\arraystretch}{1.2}
\label{firstpage}
\pagerange{\pageref{firstpage}--\pageref{lastpage}}
\maketitle

\begin{abstract}
Generally, merger likelihood increases in denser environments; however, the large relative velocities at the centres of dense clusters are expected to reduce the likelihood of mergers for satellite galaxies. Tidal features probe the recent merger histories of galaxies. The Vera C. Rubin Observatory's Legacy Survey of Space and Time (LSST) will produce an unprecedented sample of tidal features around millions of galaxies. We use LSST-like mock observations of galaxies at $z\sim0$ from the \textsc{eagle}, \textsc{IllustrisTNG} and \textsc{Magneticum Pathfinder} cosmological-hydrodynamical simulations to predict the occurrence rates of tidal features around satellite galaxies across group and cluster environments in the velocity-radius projected phase-space diagram to investigate the impact of these environments on tidal feature occurrence. We find that ancient infallers in the projected phase-space exhibit a decreasing tidal feature fraction with increasing halo mass, whereas recent infallers in the projected phase-space show unchanging tidal feature fractions with halo mass. Our results show, for the first time in cosmological simulations, a suppression of tidal feature fractions in the central regions of galaxy clusters, indicating a reduced merger rate due to higher cluster-centric velocities and lower galaxy total masses in the cluster centres. Using a toy model, we show that the presence of more tidal features in the recent infaller zone and cluster outskirts suggests that tidal features occur in interactions within infalling groups and dissipate by the time they are ancient infallers, indicating a $\lesssim3\pm2$ Gyr survival time of tidal features within clusters.
\end{abstract}


%%%%%%%%%%%%%%%%%%%%%%%%%%%%%%%%%%%%%%%%%%%%%%%%%%

%%%%%%%%%%%%%%%%% BODY OF PAPER %%%%%%%%%%%%%%%%%%

\section{Introduction}

In hierarchical structure formation-based models of the Universe, mergers play an important role in galaxy evolution \citep[e.g.][]{pressFormationGalaxiesClusters1974, fallFormationRotationDisc1980, rydenGALAXYFORMATIONGRAVITATIONAL1987, vandenboschUniversalMassAccretion2002, agertzFormationDiscGalaxies2011}. Mergers transform galaxies by contributing stars, gas and dark matter, changing galaxy dynamics, triggering star formation and active galactic nuclei (AGN) feedback \citep[e.g.][]{ostrikerGalaxyFormationIntergalactic1981,barnesEncountersDiskHalo1988,hernquistOriginKinematicSubsystems1991,dimatteoEnergyInputQuasars2005,springelModellingFeedbackStars2005,duboisHORIZONAGNSimulationMorphological2016,martinLimitedRoleGalaxy2017,martinRoleMergersDriving2018,davisonEAGLEsViewEx2020,martinRoleMergersInteractions2021,remusAccretedNotAccreted2022,cannarozzoContributionSituEx2023,rutherfordSAMIGalaxySurvey2024}.

There remain many detailed, untested predictions from simulations regarding the contribution of mergers to galaxy evolution that require further constraints from observations. For example, the cosmological evolution of the galaxy merger rate and merger ratios as a function of environment and redshift \citep[e.g.][]{hopkinsMERGERSLCDMUNCERTAINTIES2010,jianEnvironmentalDependenceGalaxy2012,rodriguez-gomezMergerRateGalaxies2015} and the contribution of mergers to the transformation of galaxies \citep[e.g.][]{bournaudGalaxyMergersVarious2005,hopkinsMergersBulgeFormation2010a,martinLimitedRoleGalaxy2017,bottrellIllustrisTNGHSCSSPImage2024}. Furthermore, the infall of galaxies into groups and clusters are predicted to contribute to the distribution of dark matter and baryons within the cluster through processes such as tidal stripping \citep[e.g.][]{ghignaDarkMatterHaloes1998,gnedinTidalEffectsClusters2003,mihosInteractionsMergersCluster2003,lokasTidalEvolutionGalaxies2020} and ram pressure stripping \citep[e.g.][]{gunnInfallMatterClusters1972,abadiRamPressureStripping1999}. 

Cosmological simulations have predicted that minor mergers (mass ratio $<1:4$) play a dominant role in the size growth \citep[][also shown analytically in \citealt{nipotiEvolutionMassiveQuiescent2025}]{naabMINORMERGERSSIZE2009} and morphological evolution \citep{naabMINORMERGERSSIZE2009,martinRoleMergersDriving2018,lagosQuantifyingImpactMergers2018,lagosConnectionMassEnvironment2018} of galaxies since $z\sim1$. Predictions regarding the environmental dependence of galaxy interaction and merger rates from cosmological simulations \citep[e.g.][]{gnedinTidalEffectsClusters2003, mihosInteractionsMergersCluster2003,omoriGalaxyMergersSubaru2023b} and semi-analytic models \citep{jianEnvironmentalDependenceGalaxy2012} indicate that the interaction rates peak at halo masses corresponding to groups and fall towards both lower halo masses (field galaxies) and higher halo masses (clusters).

Observationally, studies have highlighted a higher-than-expected rate of post-merger features in the cluster environment, sometimes comparable to the occurrence rate in the field \citep{sheenPOSTMERGERSIGNATURESREDSEQUENCE2012,ohKYDISCGalaxyMorphology2018}. The rates of these merger features are attributed to galaxies undergoing mergers in low-velocity dispersion environments such as cluster outskirts and galaxy groups, and then falling into the cluster environment \citep[e.g.][]{sheenPOSTMERGERSIGNATURESREDSEQUENCE2012,kimDistributionMergingPostmerger2024}. Works such as \citet{rheePhasespaceAnalysisGroup2017} and \citet{pasqualiPhysicalPropertiesSDSS2019} have used simulations to relate cluster-centric positions and velocities of galaxies in the velocity-radius phase-space to the expected infall time of the galaxy into a cluster, further contextualising trends in the projected phase-space diagram, and this phase-space has been used to study the occurrence of merger-related features. \citet{kimDistributionMergingPostmerger2024} found that ongoing merger-related tidal features were more likely in groups in the galaxy outskirts, while post-merger tidal features were present throughout the projected phase-space.

To test these more sensitive predictions of cosmological simulations, we need to study galaxy mergers observationally. One of the methods to do this is by studying tidal features around galaxies. Tidal features are visible signatures of ongoing or past mergers within galaxies, in the form of diffuse, non-uniform distributions of stellar light extending out from a galaxy. They have a range of morphologies, including tails/streams, shells, asymmetric haloes and double nuclei \citep[e.g.][]{zwickyMultipleGalaxies1956,malinCatalogEllipticalGalaxies1983,bilekCensusClassificationLowsurfacebrightness2020,solaCharacterizationLowSurface2022,desmonsGalaxyMassAssembly2023,valenzuelaStreamComeTrue2024}. Observations and simulations suggest that these features have lifetimes of $\sim0.7$ Gyr to $4$ Gyr \citep[e.g.][]{jiLifetimeMergerFeatures2014,KarademirOuterStellarHalos2019,mancillasProbingMergerHistory2019,yoonFrequencyTidalFeatures2020,huangMassiveEarlyTypeGalaxies2022}. Tidal features are an important probe of the recent merger histories of galaxies, however, feature detection requires very deep images. \citet{martinPreparingLowSurface2022} found that at surface brightness limits of $\mu_r\sim30$ to $31$ mag/arcsec$^{2}$ ($3\sigma$, $10^{\prime\prime}\times10^{\prime\prime}$), observations would resolve $\sim80\%$ of the flux belonging to tidal features around a Milky Way mass galaxy out to $z\sim0.2$.

To use observations of tidal features around galaxies to test the predictions made by cosmological simulations, it is necessary to investigate the occurrence of these features in simulations. N-body and hydrodynamical simulations have been used to constrain how tidal features trace the angular momentum and mass ratios of the mergers that created them. Tail and stream-like tidal features form in simulations due to high angular momentum mergers of galaxies \citep[e.g.][]{toomreGalacticBridgesTails1972,hendelTidalDebrisMorphology2015,KarademirOuterStellarHalos2019,valenzuelaStreamComeTrue2024}, whereas shells occur in simulations due to radial mergers \citep[e.g.][]{amoriscoFeathersBifurcationsShells2015,hendelTidalDebrisMorphology2015,popFormationIncidenceShell2018,valenzuelaStreamComeTrue2024}.

\citet{valenzuelaStreamComeTrue2024} and \citet{khalidCharacterizingTidalFeatures2024} found that the occurrence of tidal features increases with increasing galaxy stellar mass in cosmological simulations. A result that has also been seen in observations \citep[e.g.][]{bilekCensusClassificationLowsurfacebrightness2020,desmonsGalaxyMassAssembly2023}. \citet{valenzuelaStreamComeTrue2024} did, however, note a tension with the tidal feature fractions as a function of stellar mass sitting below the expected relation from the MATLAS survey \citep{bilekCensusClassificationLowsurfacebrightness2020}.

% POTENTIALLY IN DISCUSSION
% and the occurrence of shells correlates with slow-rotator galaxies, highlighting the importance of tidal features in linking mergers to present galaxy morphology \citep[also seen in observations][]{yoonEvidenceImpactGalaxy2022,rutherfordSAMIGalaxySurvey2024}.

With the upcoming Vera C. Rubin Observatory's Legacy Survey of Space and Time \citep[LSST;][]{ivezicLSSTScienceDrivers2019a,robertsonGalaxyFormationEvolution2019,broughVeraRubinObservatory2020}, it will be possible to study tidal features around millions of galaxies \citep{martinPreparingLowSurface2022}, allowing for the most robust statistical survey of tidal features to date. In \citet{khalidCharacterizingTidalFeatures2024}, we undertook an observationally-motivated analysis of tidal features around galaxies in cosmological simulations. To do this we produced a set of mock observations at the predicted 10-year LSST surface brightness limits \footnote{Peter Yaochim, \href{https://smtn-016.lsst.io/}{https://smtn-016.lsst.io/}} and visually identified the tidal features present around galaxies from four state-of-the-art cosmological-hydrodynamical simulations \textsc{NewHorizon} \citep{duboisIntroducingNEWHORIZONSimulation2021}, \textsc{EAGLE} \citep{schayeEAGLEProjectSimulating2015, crainEAGLESimulationsGalaxy2015}, \textsc{IllustrisTNG} \citep{pillepichFirstResultsIllustristng2018,springelFirstResultsIllustrisTNG2018, nelsonFirstResultsIllustrisTNG2018, marinacciFirstResultsIllustrisTNG2018, naimanFirstResultsIllustrisTNG2018} and \textsc{Magneticum Pathfinder} \citep{tekluConnectingAngularMomentum2015,dolagEncyclopediaMagneticumScaling2025}.

\citet{khalidCharacterizingTidalFeatures2024} also made predictions regarding the occurrence of tidal features as a function of the host galaxy's stellar mass and host halo mass. We found that the occurrence of tidal features varied as a function of group/cluster halo mass even after accounting for the stellar mass of the galaxies, peaking at $\log_{10}(M_{\scriptstyle\mathrm{200,\:crit}}/$M$_{\scriptstyle\odot})\sim12.7$ and declining after this. We found that this peak was driven primarily by the satellite galaxy populations and that central galaxies followed a different trend, where tidal feature fractions increased with increasing halo mass. However, further investigation is required to see if the occurrence of tidal features around simulated galaxies in the cluster environment is consistent with the observational results, which suggest the mergers occur in lower-velocity encounters outside the cluster and that the features last during infall.

% It is necessary to validate whether these changes in tidal feature occurrence rates are tracing the measured peaks in the galaxy merger rates in the group regime \citep{gnedinTidalEffectsClusters2003,jianEnvironmentalDependenceGalaxy2012} or if the driving factor is the lifetime of tidal features \citep[e.g.][]{mancillasProbingMergerHistory2019} and any potential variance of this with the environment.

In this paper, we will use the projected phase-space to determine how tidal feature occurrence rates relate to the relative positions and velocities of galaxies to the cluster centre in our LSST-like mock observations. This will enable us to further investigate the role played by the environment in the occurrence of tidal features around simulated galaxies.

In Section \ref{sec:data_methods}, we briefly describe the simulations (Section \ref{subsec:simulations}), the catalogue of tidal features around galaxies (Section \ref{subsec:tf_catalogue}) and how we use the velocity-radius phase-space to analyse the occurrence of tidal features in groups and clusters of galaxies (Section \ref{subsec:meth_vel_rad}), by separating galaxies into ancient and recent infallers based on the \citet{pasqualiPhysicalPropertiesSDSS2019} definitions. In Section \ref{section:results} we present our results and discuss them in Section \ref{sec:discussion}. We draw our conclusions in Section \ref{sec:conclusions}. We use the native cosmology from each simulation for calculating distances between particles and creating our mock images; these are given in Table \ref{tab:simulations}. For distances we use a `c' prefix to denote comoving distances and a `p' prefix to denote proper distances, e.g. `ckpc' is comoving kiloparsecs and `pkpc' is proper kiloparsecs.

\section{Data and Methods}
\label{sec:data_methods}

For this work, we use the catalogue of visually identified tidal features around LSST-like mock images of galaxies from three\footnote{\textsc{NewHorizon} lacks cluster mass halos.} state-of-the-art cosmological-hydrodynamical simulations, \textsc{eagle} \textit{RefL100N1504}, \textsc{IllustrisTNG} \textit{L75N1820}, \textsc{Magneticum Pathfinder} \textit{Box4-uhr} (from here on EAGLE, TNG and Magneticum). The specifications of these simulations are summarised in Table~\ref{tab:simulations}. The production of the mock images and the construction of the catalogue are presented by \citet{khalidCharacterizingTidalFeatures2024}. The mock images span galaxies with stellar masses ranging from $9.5<\log_{10}(M_{\scriptstyle\star\mathrm{,\:30\:pkpc}}/$M$_{\scriptstyle\odot})<11.8$, where $M_{\scriptstyle\star\mathrm{,\:30\:pkpc}}$ is the total amount of stellar mass contained within a 30 pkpc radius spherical aperture centred on the galaxy centre of potential. \citet{schayeEAGLEProjectSimulating2015} showed that the 30 pkpc radius spherical aperture provides results for the stellar mass halo mass relation that are very similar to the 2D Petrosian aperture \citep{petrosianSurfaceBrightnessEvolution1976} stellar masses often used in observations, therefore, for our observationally motivated analysis we use this stellar mass.

This work aims to investigate further how the phase-space location of a galaxy within a group or a cluster relates to the frequency of tidal feature occurrence. In particular, we explore how the occurrence of tidal features differs when transitioning from halo masses corresponding to a low-mass group environment ($12<\log_{10}(M_{\scriptstyle\mathrm{200,\:crit}}/$M$_{\scriptstyle\odot})<13$) to halo masses corresponding to high-mass groups ($13<\log_{10}(M_{\scriptstyle\mathrm{200,\:crit}}/$M$_{\scriptstyle\odot})<14$) and galaxy clusters ($14<\log_{10}(M_{\scriptstyle\mathrm{200,\:crit}}/$M$_{\scriptstyle\odot})<14.7$). Here, halo masses are measured using $M_{\scriptstyle\mathrm{200,\:crit}}$, which is the total mass contained within the radius at which the mean contained density of a halo has dropped to 200 times the critical density of the universe ($R_{\scriptstyle\mathrm{200,\:crit}}$).

\subsection{Simulations}
\label{subsec:simulations}
\begin{table*}
	\centering
	\caption{Summary of the properties of the three cosmological-hydrodynamical simulations. From left to right, the columns are the simulation name, the cosmology used to determine the initial conditions, the simulation volume, the redshift corresponding to the snapshot used in this study, the dark matter mass resolution, the initial baryonic mass resolution, the mean stellar mass resolution in the simulation snapshot, the size of the catalogue of classified galaxies, the number of satellite galaxies and the fraction of satellite galaxies that have tidal features.}
	\label{tab:simulations}
	\begin{tabular}{lccccccccr} 
		\hline
		Simulation & Cosmology & Volume [cMpc$^3$] & $z$ & $m_{\scriptstyle\mathrm{DM}}$ [M$_{\scriptstyle\odot}$] & $m_{\scriptstyle\mathrm{gas}}$ [M$_{\scriptstyle\odot}$] & $m_{\scriptstyle\star}$ [M$_{\scriptstyle\odot}$] & $N$ & $N_{\scriptstyle\mathrm{Satellite}}$ & $f_{\scriptstyle\mathrm{Tidal,\:satellite}}$\\
		\hline
		EAGLE & Planck13 & $100^3$ & 0.052 & $9.7\times10^6$ & $1.81\times10^6$ & $1.09\times10^6$ & 1983 & 829 & $0.22^{+0.02}_{-0.01}$\\
		TNG & Planck15 & $110.7^3$ & 0.05 & $7.5\times10^6$ & $1.4\times10^6$ & $1.07\times10^6$ & 1826 & 699 & $0.25\pm0.02$\\
		Magneticum & WMAP7 & $68^3$ & 0.066 & $5.11\times10^7$ & $1.04\times10^7$ & $1.85\times10^6$ & 1989 & 747 & $0.22\pm0.02$\\
		\hline
	\end{tabular}
\end{table*}

\citet{khalidCharacterizingTidalFeatures2024} showed that EAGLE, TNG and Magneticum all exhibited similar behaviour regarding the occurrence of tidal features as a function of galaxy stellar mass and halo mass, suggesting that the occurrence of tidal features was independent of the different subgrid galaxy formation models applied by each simulation. While the intrinsic properties of the tidal features. For example, their colour differs between simulations (Khalid et al. in prep), as we are focused only on their detection/occurrence rates, we expect the three simulations to behave similarly and allow for a combined sample of clusters to enhance the statistical robustness of the study. In this section, we briefly describe each simulation and highlight the ways in which differences between simulations can lead to differences in the occurrence of tidal features.

The \textsc{eagle} (Evolution and Assembly of GaLaxies and their Environments) project is a large set of cosmological hydrodynamical simulations. \textsc{eagle} contains simulations of cubic volume $12^3$, $25^3$, $50^3$ and $100^3$ cMpc$^{3}$. The data used in this study is taken from the $z=0.052$ snapshot of the reference model (\textit{RefL100N1504}), which has a volume of $100^3\text{ cMpc})^3$. The reference model is the simulation run with standard parameters and physics as described in \citet{schayeEAGLEProjectSimulating2015} and \citet{crainEAGLESimulationsGalaxy2015}. The simulation uses the cosmological parameters advocated by the Planck 2013 results \citep[][]{adePlanck2013Results2014}. The dark matter particle mass is $m_{\rm DM}=9.70 \times 10^{6} \text{ M}_{\scriptstyle \odot}$, the initial mass of the gas particles is $m_\mathrm{gas}=1.81\times10^6$ M$_{\scriptstyle\odot}$ and the mean mass of stellar particles at $z=0.052$ is $m_{\scriptstyle\star}=1.09\times10^{6} \text{ M}_{\scriptstyle \odot}$. The simulation was calibrated to match the galaxy stellar mass function (GSMF) and the relationship between black hole mass and galaxy stellar mass at $z\sim0$ \citep[details in][]{schayeEAGLEProjectSimulating2015}.

\textsc{IllustrisTNG} is a suite of simulations, with the following box volumes: $51.7^3$, $110.7^3$ and $302.6^3\text{ cMpc}^3$. We use the $z=0.05$ snapshot of the simulation \textit{L75N1820TNG}, which models the physics of dark matter and baryons in a $110.7^3\text{ cMpc}^3$ box \citep{marinacciFirstResultsIllustrisTNG2018,naimanFirstResultsIllustrisTNG2018,nelsonFirstResultsIllustrisTNG2018,springelFirstResultsIllustrisTNG2018,pillepichFirstResultsIllustristng2018}. The galaxy formation models used are described in \citet{weinbergerSimulatingGalaxyFormation2017} and \citet{pillepichSimulatingGalaxyFormation2018}. This simulation adopts the $\Lambda$CDM model fit by the Planck 2015 results \citep[]{adePlanck2015Results2016}. The dark matter resolution for the simulation is $m_{\rm DM}=7.5\times10^6\text{ M}_{\scriptstyle \odot}$, the initial mass of gas particles is $m_{\scriptstyle\mathrm{gas}}=1.4\times10^6\text{ M}_{\scriptstyle \odot}$ and the average mass of stellar particles at $z=0.05$ is $m_{\scriptstyle\star}=1.07\times10^6$ M$_{\scriptstyle\odot}$. \textsc{IllustrisTNG} is calibrated to match the star formation rate density, the GSMF and the stellar-to-halo mass relation, the total gas mass contained within the virial radius of massive groups, the stellar mass – stellar size and the black hole – galaxy mass relations all at $z=0$, in addition to the overall shape of the cosmic star formation rate density at $z\lesssim10$ \citep{pillepichFirstResultsIllustristng2018}.

\textsc{Magneticum Pathfinder} simulations are a suite of cosmological hydrodynamical simulations, ranging in box size from $25.6^3$ to 3818$^3$ cMpc$^3$. We use the $z=0.066$ snapshot from the \textsc{Box4-uhr} simulation, which models the physics of dark matter and baryons in a $68^3$ cMpc$^3$ box. All simulations are performed with an updated version of the SPH code \textsc{GADGET-3} \citep{springelSimulationsFormationEvolution2005}. The galaxy formation model for \textsc{Magneticum Pathfinder} is described in \citet{tekluConnectingAngularMomentum2015} and \citet{dolagEncyclopediaMagneticumScaling2025}. We use the \textit{Box4-uhr} run, which adopts a WMAP-7 fit cosmology \citep{komatsuSevenyearWilkinsonMicrowave2011}. The mass resolution for dark matter particles is $m_{\scriptstyle\mathrm{DM}}=5.11\times10^7$ M$_{\scriptstyle\odot}$, the initial mass of gas particles is $m_{\scriptstyle\mathrm{gas}}=1.04\times10^7$ M$_{\scriptstyle\odot}$ and the average mass of stellar particles at $z=0.066$ is $m_{\scriptstyle\star}=1.85\times10^6$ M$_{\scriptstyle\odot}$. The Magneticum simulations are calibrated to match the intracluster gas content of galaxy clusters, rather than matching any individual galaxies' characteristics \citep{popessoAverageXrayProperties2024,dolagEncyclopediaMagneticumScaling2025}.

The differences in simulation dark matter resolution and subgrid physics models could impact the formation and lifetimes of tidal features. Lower dark matter resolution could result in artificially cored halos, leading to stellar stripping becoming too efficient \citep[][]{martinStellarStrippingEfficiencies2024}, potentially reducing the lifetime of tidal features within clusters \citep{jiLifetimeMergerFeatures2014}. Furthermore, numerical heating effects driven by lower resolution dark matter particles may inhibit the formation of a dynamically cold disk \citep[e.g.][]{ludlowSpuriousHeatingStellar2021,ludlowSpuriousHeatingStellar2023}. Given that tidal features formed from disk galaxies are more prominent and longer-lived than those formed from elliptical galaxies \citep[e.g.][]{feldmannTidalDebrisElliptical2008,souchayTidesAstronomyAstrophysics2013}, the resolution of cold disk orbits will impact the observability and lifetime of the tidal features produced. However, given the similar dark matter resolutions of the simulations and the similar occurrence of tidal features with galaxy stellar mass and halo mass \citep{khalidCharacterizingTidalFeatures2024}, differences in subgrid physics models could play a larger role by producing differences between the structural and dynamical properties of the galaxies across simulations \citep[e.g.][]{vandesandeSAMIGalaxySurvey2019}. The different sizes, concentrations and orbital anisotropies change the detectability and lifetimes of the tidal features produced \citep[e.g.][]{feldmannTidalDebrisElliptical2008,penarrubiaImpactDarkMatter2010,souchayTidesAstronomyAstrophysics2013,martinStellarStrippingEfficiencies2024}.

\subsection{Tidal feature catalogue and sample}
\label{subsec:tf_catalogue}

A catalogue identifying tidal features and their morphologies in LSST-like mock images was constructed for a subsample of galaxies from each simulation in \citet{khalidCharacterizingTidalFeatures2024}. The sample was constructed by randomly selecting galaxies from each simulation with $M_{\star\mathrm{,\: 30\: pkpc}}>10^{9.5}$ M$_{\scriptstyle \odot}$. In Section \ref{subsubsec:mock_images}, we describe the production of the LSST-like mock images, in Section \ref{subsubsec:tf_classification}, we describe how the tidal feature catalogue was constructed and lastly in \ref{subsubsec:completeness} we measure how complete our sample of satellites is for each group/cluster in our tidal feature catalogue.

\subsubsection{Mock images}
\label{subsubsec:mock_images}

We briefly describe here the LSST-like mock images constructed from the simulated galaxies; further details are given in \citet{martinPreparingLowSurface2022} and \citet{khalidCharacterizingTidalFeatures2024}. The mock images are produced using a similar method to \citet{martinPreparingLowSurface2022}. We extract all the particles within a 1 pMpc side-length cube of the galaxy that are also within the same Friends-Of-Friends (FOF) assigned group. We place our simulated galaxy at $z=0.025$, at a comoving distance of $\sim109$ cMpc. We assign a spectral energy distribution (SED) to each stellar particle using simple stellar population models \citep{bruzualStellarPopulationSynthesis2003} interpolated to the age and metallicity of the stellar particle. Assuming a Chabrier initial mass function \citep{chabrierGalacticStellarSubstellar2003}, the \textit{g}, \textit{r}, and \textit{i} flux is calculated accounting for the galaxy redshift and the LSST bandpass functions \citep{olivierOpticalDesignLSST2008a}. Reddening from dust attenuation is not accounted for in our images. The presence of gas and dust will reduce the luminosity and, therefore, the detectability of the tidal features. However, this does not significantly impact the detection and colours of tidal features, which are often found in the dust-poor outskirts of the galaxy \citep[][]{martinPreparingLowSurface2022}.

Once the particle density drops to below a few particles per 0.2 arcsecond pixel, the \textsc{smooth3d}\footnote{\hyperlink{https://github.com/garrethmartin/smooth3d}{https://github.com/garrethmartin/smooth3d}} algorithm is applied. Each particle in the underdense region is redistributed into normally distributed lower mass elements centred around the original particle with a standard deviation equivalent to the $5^\mathrm{th}$ nearest neighbour distance \citep[][see also \citealt{merrittMissingOutskirtsProblem2020}]{martinPreparingLowSurface2022}. This smoothing prevents unrealistic variance in the image flux.

The mock images are produced by collapsing the image along the \textit{z} axis of the simulation box (i.e. \textit{x-y} projection), and binning the particles to produce the required $0.2^{\prime\prime}\times0.2^{\prime\prime}$ pixel size. This image is then convolved with the Hyper-Suprime Cam point spread function as measured by \citep[][]{montesBuildupIntraclusterLight2021} rescaled to the LSST pixel size, to model the dispersion of light due to seeing.

Lastly, we add a Gaussian background noise following the observationally measured empirical relation from \citet{romanGalacticCirriDeep2020}. The standard deviation of the Gaussian background is given by:
\begin{equation}
    \sigma_\mathrm{noise} = \frac{10^{-0.4\mu^\mathrm{lim}_\mathrm{band}(n\sigma,\Omega\times\Omega)}\mathrm{pix}\times\Omega}{n}.
\end{equation}

Here $\Omega$ is the side-length in arcseconds of one side of the square over which the surface brightness limit is computed, $=10^{\prime\prime}$, pix is the pixel scale in arcseconds per pixel, $=0.2$ arcsec/pixel, $n$ is the number of Gaussian standard deviations for the surface brightness limits, $=3$ and $\mu_{\rm band}^{\rm lim}$ is the $3\sigma$, $10^{\prime\prime}\times10^{\prime\prime}$ limiting surface brightness modelled for a particular LSST-photometric band ($\mu_{g}^{\mathrm{lim}}=30.3$ mag/arcsec$^{2}$, $\mu_{r}^{\mathrm{lim}}=30.3$ mag/arcsec$^{2}$, $\mu_{i}^{\mathrm{lim}}=29.7$ mag/arcsec$^{2}$). 

% \footnote{Peter Yaochim, \href{https://smtn-016.lsst.io/}{https://smtn-016.lsst.io/}}
\subsubsection{Tidal feature catalogue}
\label{subsubsec:tf_classification}

The galaxies were visually classified by the first author in \citet{khalidCharacterizingTidalFeatures2024} to catalogue the presence of any tidal features around any of the galaxies and classify them into the following categories:
\begin{itemize}
    \item \textbf{Streams/Tails}: Prominent, elongated structures orbiting or expelled from the host galaxy.
    \item \textbf{Shells}: Concentric radial arcs or ring-like structures around a galaxy.
    \item \textbf{Plumes or Asymmetric Stellar haloes}: Diffuse features in the outskirts of the host galaxy, lacking well-defined structures like stellar streams or tails.
    \item \textbf{Double nuclei}: Two clearly separated galaxies within the mock image where merging is evident through the presence of tidal features.
\end{itemize}

Furthermore, for each tidal feature classification, the catalogue provides a confidence level ranging from zero to three.
\begin{enumerate}
    \item[(0)] No tidal feature detected.
    \item[(1)] Hint of tidal feature detected, classification difficult.
    \item[(2)] Even chance of correct classification of tidal feature presence and/or morphology.
    \item[(3)] High likelihood of the tidal feature being present and morphology being obvious.
\end{enumerate}

To minimize misclassified tidal features, we use the confidence level 3 tidal features. The tidal feature fractions for the sample of satellite galaxies given this classification level are shown in Table \ref{tab:simulations}. In this analysis, we estimate uncertainties on our feature fractions using the $1\sigma\simeq0.683$ binomial confidence levels estimated using a Bayesian beta distribution generator for binomial confidence intervals described in detail in \citet{cameronEstimationConfidenceIntervals2011}. This approach is robust to small sample sizes, which can be the case for subsamples of our satellite populations within the velocity-radius-phase-space.

\begin{table}
    \centering
    \caption{The number of haloes and satellite subhaloes in our sample for each $M_{\scriptstyle\mathrm{200,\:crit}}$ bin for each simulation.}
    \begin{tabular}{l|c|c|r}
        \hline
         Simulation & & Number of haloes &\\
         \hline
         $M_{\scriptstyle\mathrm{200,\:crit}}$ bins& $[10^{12},\:10^{13}]$ M$_{\scriptstyle\odot}$ & $[10^{13},\:10^{14}]$ M$_{\scriptstyle\odot}$ & $\geq10^{14}$ M$_{\scriptstyle\odot}$ \\
         \hline
         EAGLE & 210 & 127 & 7\\
         TNG & 193 &  122 & 14\\
         Magneticum & 141 & 38 & 3\\
         \hline
         & & Number of satellites & \\
        \hline
         EAGLE & 25 & 331 & 205\\
         TNG & 211 &  290 & 157\\
         Magneticum & 231 & 306 & 109\\
         \hline
    \end{tabular}
    \label{tab:numb_subhalo_halo}
\end{table}

\begin{figure}
    \centering
    \includegraphics[width=\linewidth]{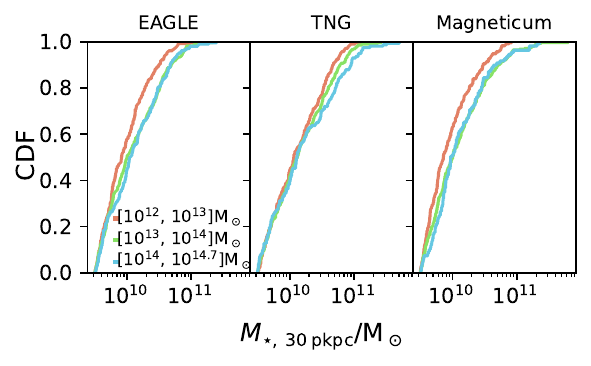}
    \caption{The cumulative distribution functions for satellite stellar mass computed for 3 different halo mass bins. The red lines show the distribution for $10^{12}<M_{\scriptstyle\mathrm{200,\:crit}}/\mathrm{M}_{\scriptstyle\odot}<10^{13}$, the green lines show the distribution for $10^{13}<M_{\scriptstyle\mathrm{200,\:crit}}/\mathrm{M}_{\scriptstyle\odot}<10^{14}$ and the blue line shows the distribution for $M_{\scriptstyle\mathrm{200,\:crit}}\geq10^{14}\mathrm{\:M}_{\scriptstyle\odot}$. Stellar mass is measured using a 30 pkpc spherical aperture. We see across all simulations that the stellar mass distributions for the intermediate and highest halo mass bins are shifted to higher stellar masses than the distributions for the lowest halo mass bin.}
    \label{fig:sample_mstar}
\end{figure}

For this work, we rely on the FOF and \textsc{SubFind} \citep[]{springelPopulatingClusterGalaxies2001,dolagSubstructuresHydrodynamicalCluster2009} identified groups (haloes) and galaxies (subhaloes). \textsc{SubFind} identifies the most massive subhalo in a halo as the central, and the less massive subhaloes are identified as satellites of this central. The sample of galaxies used in this work is summarised in Table~\ref{tab:simulations}. Table~\ref{tab:numb_subhalo_halo}, gives the number of haloes (groups/clusters) and the number of satellite subhaloes (satellite galaxies) in our samples for halo mass bins, $M_{\scriptstyle\mathrm{200,\:crit}}=[10^{12},\:10^{13},\:10^{14},\:10^{15}]$ M$_{\scriptstyle\odot}$. The simulations have most of their subhaloes residing in haloes in the intermediate $10^{13}\leq M_{\scriptstyle\mathrm{200,\:crit}}/\mathrm{M}_{\scriptstyle\odot}<10^{14}$ bin. The TNG sample contains the largest number of $M_{\scriptstyle\mathrm{200,\:crit}}\geq10^{14}$ M$_{\scriptstyle\odot}$ haloes. TNG has a similar number of haloes to EAGLE in the low and intermediate halo mass. Magneticum has fewer haloes across all halo mass bins. This is not surprising as the Magneticum simulation has a smaller volume than the other two simulations.

Fig.~\ref{fig:sample_mstar} shows the stellar mass distributions for the satellite galaxies in each simulation across the three halo mass bins ($10^{12}<M_{\scriptstyle\mathrm{200,\: crit}}/\mathrm{M}_{\scriptstyle\odot}<10^{13}$, $10^{13}<M_{\scriptstyle\mathrm{200,\:crit}}/\mathrm{M}_{\scriptstyle\odot}<10^{14}$ and $10^{14}<M_{\scriptstyle\mathrm{200,\:crit}}/\mathrm{M}_{\scriptstyle\odot}<10^{14.7}$). The lowest halo mass corresponds to low mass groups, whereas the middle and highest halo mass bins correspond to massive groups and cluster environments, respectively. For all three simulations, the higher two halo mass bins have satellite stellar mass distributions that are shifted to higher stellar masses when compared to the lowest halo mass bin. The shift to higher stellar masses is expected as a consequence of hierarchical structure formation, given that more massive groups/clusters will form through the mergers of more massive haloes than smaller groups/clusters \citep[e.g.][]{pressFormationGalaxiesClusters1974,whiteCoreCondensationHeavy1978}. Given that we know the identified tidal feature fraction increases with increasing stellar mass \citep[e.g.][]{khalidCharacterizingTidalFeatures2024}, it will be important to consider this potential source of bias in our analysis.

\subsubsection{Catalogue completeness}
\label{subsubsec:completeness}

\begin{figure}
    \centering
    \includegraphics[width=\linewidth]{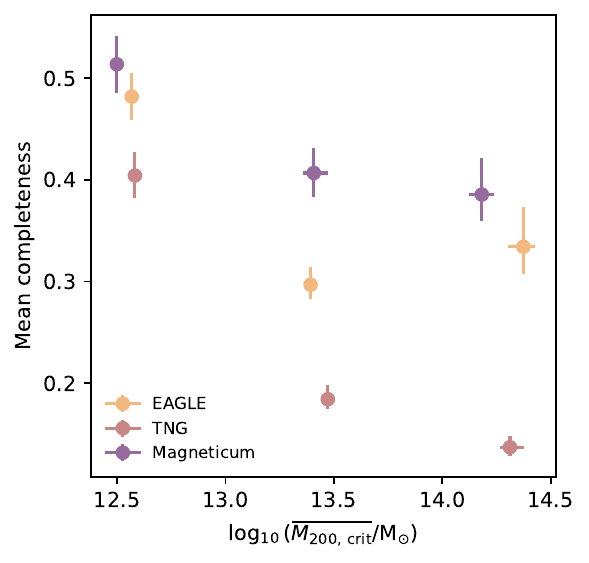}
    \caption{The mean completeness of each FOF group in our sample, as a function of the mean halo mass for EAGLE (orange), TNG (pink) and Magneticum (purple). The points and uncertainties show the median and $1\sigma$ confidence levels for each value measured by bootstrapping. The completeness decreases with increasing halo mass. The sample for Magneticum is the most complete, followed by EAGLE, then TNG.}
    \label{fig:completeness}
\end{figure}

% In \citet{khalidCharacterizingTidalFeatures2024}, the selection of galaxies for the visual classification of tidal features was done by randomly sampling galaxies above $M_{\star}\geq10^{9.5}$ M$_\odot$ ($M_\star$ is the total stellar mass as calculated from the \textsc{SubFind} identified galaxy) as this would enable us to see how tidal feature fractions varied with stellar mass. This results in a catalogue where not every galaxy in each of the groups and clusters in our sample has been selected.

While the \citet{khalidCharacterizingTidalFeatures2024} catalogue did sample galaxies across a range of galaxy stellar masses and parent halo masses, the group and cluster satellites sampled are a fraction of the total number of satellites with $M_{\star,\:\mathrm{30\:pkpc}}>10^{9.5}\:\mathrm{M}_\odot$ in each simulation. For this analysis, it is important to consider how complete the sample of satellites in each parent halo is for each simulation. For $10^{12}<M_\mathrm{200,\:crit}/\mathrm{M}_\odot<10^{13}$, we sample 18 per cent of the parent halos in EAGLE, 11 per cent in TNG and 41 per cent in Magneticum, for $10^{13}<M_\mathrm{200,\:crit}/\mathrm{M}_\odot<10^{14}$ these numbers increase to 83 per cent, 69 per cent and 100 per cent respectively, and lastly for $10^{14}<M_\mathrm{200,\:crit}/\mathrm{M}_\odot<10^{14.7}$ they increase to 100 percent for all simulations. Fig.~\ref{fig:completeness} shows the mean completeness of each group in the catalogue for each simulation as a function of halo mass. This is computed by taking the number of satellites in each group in our catalogue and dividing this number by the number of satellites assigned by FOF+\textsc{SubFind} to that group. As we only consider galaxies within $3\:R_\mathrm{200,\:crit}$ of the group/cluster centre in \textit{x}-\textit{y} projected distance in this work (see Section \ref{subsec:meth_vel_rad}), we also only compute the completeness within this region. Magneticum and EAGLE have a mean completeness of $\sim0.5$ for the $12.5<\log_{10}(M_\mathrm{200,\:crit}/\mathrm{M}_\odot)<13$ halo mass bin, while TNG has a mean completeness of $0.40\pm0.02$. The mean completeness for Magneticum remains at $\sim0.4$ for the $13<\log_{10}(M_\mathrm{200,\:crit}/\mathrm{M}_\odot)<14$ and $14<\log_{10}(M_\mathrm{200,\:crit}/\mathrm{M}_\odot)<14.7$ halo mass bins; EAGLE has a mean completeness of $0.30\pm0.01$ and $0.33^{+0.04}_{-0.03}$ for these halo mass bins and TNG has a mean completeness of $0.18\pm0.01$ and $0.14\pm0.01$, respectively. 

The higher completeness of Magneticum is unsurprising, as it has a smaller simulation box. As such, the 747 satellites in our catalogue comprise a larger proportion of the total number of satellite galaxies in each group and cluster in our sample. The lower completeness of TNG is a result of the smaller sample of galaxies from this simulation in our catalogue, due to it being disproportionately impacted by the removal of compact point-like objects, which were not realistic galaxies. This results in us sampling a smaller proportion of the satellites in the groups and clusters in this simulation.

In general, the low levels of completeness could suggest that the predictions from individual simulations, particularly TNG in the cluster regime, could suffer from having too small a sample to be representative of the entire population. However, we are assisted somewhat by the random nature of the sampling, reducing the likelihood of any systematic bias. We explore the impact that incompleteness might have further in Appendix \ref{app:completeness} and also quantify the number of satellites \textsc{SubFind} assigns to each of the groups and clusters (i.e. group/cluster richness) in Appendix \ref{app:richness}.

% % \subsubsection{Uncertainties on fractions}
% We estimate uncertainties on our feature fractions using the $1\sigma\simeq0.683$ binomial confidence levels estimated using a Bayesian beta distribution generator for binomial confidence intervals described in detail in \citet{cameronEstimationConfidenceIntervals2011}. This approach is robust to small sample sizes, which can be the case for subsamples of our satellite populations within the velocity-radius phase-space.

\subsection{Velocity and radius phase-space}
\label{subsec:meth_vel_rad}

To analyse the behaviour of tidal features in groups and clusters of galaxies, we follow the regions in the velocity-radius projected phase-space defined by \citet{pasqualiPhysicalPropertiesSDSS2019}. These regions allow for predictions regarding the infall times that are directly comparable to observations and therefore are an ideal choice for our aim of providing observationally-testable predictions. The regions are defined out to 1 $R_{\scriptstyle\mathrm{200,\:crit}}$ \footnote{This approximates the virial radius at $z\sim0$ \citep{bryanStatisticalPropertiesXRay1998}.}, using the following equation:

\begin{equation}
    \label{eq:v_r_pasquali}
    \left|\frac{\Delta V_\mathrm{LOS}}{\sigma_\mathrm{LOS}}\right| = a\left(\frac{R_{\scriptstyle\mathrm{proj}}}{R_{\scriptstyle\mathrm{200,\:crit}}}\right)^2 + b\left(\frac{R_{\scriptstyle\mathrm{proj}}}{R_{\scriptstyle\mathrm{200,\:crit}}}\right) + c,
\end{equation}

where $\Delta V$ is the line-of-sight velocity relative to the cluster centre, and $\sigma_\mathrm{LOS}$ is the line-of-sight velocity dispersion of the group at $R_\mathrm{200,\:crit}$. $R_\mathrm{proj}$ is the projected distance from the cluster centre. We use the intrinsic \textit{x}-\textit{y} projection from each simulation to measure the projected distances and line-of-sight velocity. The coefficients, $a$, $b$ and $c$ are expressed as a function of a positive integer $p$, between 1 and 7:
\begin{align}
    a &= 0.022p^3-0.512p^2+3.287p-2.786,\label{eq:a}\\
    b &= 0.184p^2-1.494p-3.5,\label{eq:b}\\
    c &= -0.108p^2 + 1.249p + 0.314.\label{eq:c}
\end{align}

Following \citet{pasqualiPhysicalPropertiesSDSS2019}, we classify galaxies within the zone defined by $p\leq2$ as ancient infallers, which \citet{pasqualiPhysicalPropertiesSDSS2019} found had an average infall time $5\pm2$ Gyr, and we classify galaxies within the zone defined by $p\geq5$ and $R_\mathrm{proj}\leq1\:R_\mathrm{200,\:crit}$ as recent infallers with an average infall time $2\pm1$ Gyr. For the intermediate zone, the average infall time from \citet{pasqualiPhysicalPropertiesSDSS2019} is $4\pm2$ Gyr. \citet{pasqualiPhysicalPropertiesSDSS2019} defined these zones based on clusters ranging in halo mass from $5.3\times10^{13}$ M$_\odot$ to $9.2\times10^{14}$ M$_\odot$, meaning their infall times provide rough estimates for our $[10^{13},\:10^{14}]$ M$_\odot$ and $[10^{14},\:10^{14.7}]$ M$_\odot$ bins. We test in Appendix \ref{app:infall_times} how the infall times for each zone compare with the infall times for EAGLE from \citet{wrightOrbitalPerspectiveStarvation2022} and find that the infall times in each zone are consistent. The region beyond $R_\mathrm{200,\:crit}$ is dominated by interlopers and therefore, \citet{pasqualiPhysicalPropertiesSDSS2019} limits their zone definitions to within $R_\mathrm{200,\:crit}$. In our work, we also examine the tidal feature fractions for \textsc{SubFind} identified satellites in the region beyond $R_\mathrm{200,\:crit}$ up to $3\:R_\mathrm{200,\:crit}$ for comparison to the recent and ancient infallers and refer to this region as the outskirts of the group/cluster. We make the radial cutoff at $R_\mathrm{proj}\leq3\:R_\mathrm{200,\:crit}$ as the satellite population beyond this radius has been observed to have star formation properties consistent with those of field galaxies \citep[e.g.][]{lewis2dFGalaxyRedshift2002,barsantiGalaxyMassAssembly2018}.

In observations of groups and clusters, we can only obtain the line-of-sight (LOS) velocity dispersion, $\sigma_{\scriptstyle\mathrm{LOS}}$. However, from the simulation we have the measured $R_{\scriptstyle\mathrm{200,\:crit}}$ and $M_{\scriptstyle\mathrm{200,\:crit}}$ and given that galaxy clusters are dominated by velocity dispersion, we can estimate $\sigma_{\scriptstyle\mathrm{200,\:crit}}$ from the virial theorem as:

% Using the true velocities and distances will reduce the spread on the average infall times computed by \citet{pasqualiPhysicalPropertiesSDSS2019}.

\begin{equation}
    \sigma_{\scriptstyle\mathrm{200,\:crit}}=\sqrt{\frac{GM_{\scriptstyle\mathrm{200,\:crit}}}{R_{\scriptstyle\mathrm{200,\:crit}}}}.
\end{equation}

Given $\sigma_\mathrm{200,\:crit}=\sqrt{\sigma_\mathrm{x,\:200\:crit}^2+\sigma_\mathrm{y,\:200\:crit}^2+\sigma_\mathrm{z,\:200\:crit}^2}$, assuming that the velocities of the cluster satellite galaxies are isotropic, then we can estimate $\sigma_\mathrm{LOS}$ as:
\begin{equation}
    \sigma_\mathrm{LOS}=\frac{\sigma_\mathrm{200,\:crit}}{\sqrt{3}}.
\end{equation}

By projecting the galaxy on the $x$-$y$ plane of the simulation box, our LOS is the $z$ axis.

\section{Results}
\label{section:results}

\begin{table}
    \centering
    \caption{Tidal feature fraction in each of the zones defined by \citet{pasqualiPhysicalPropertiesSDSS2019} across three halo mass bins, $M_{\scriptstyle\mathrm{200,\:crit}}=[10^{12},\:10^{13},\:10^{14},\:10^{14.7}]$ M$_{\scriptstyle\odot}$. The number of satellites in each zone and bin is given within parentheses. From left to right, the columns give the zone, the results for EAGLE, TNG and Magneticum. The uncertainties on the fractions give the $1\sigma$ binomial confidence intervals. Blank spaces in the table indicate that there are no galaxies in that zone.}
    \begin{tabular}{l|l|l|l}
    \hline
        Zone [$p$] & EAGLE & TNG & Magneticum \\
    \hline
          & \multicolumn{3}{c}{$10^{12}< M_\mathrm{200,\:crit}/$M$_\odot<10^{13}$}\\
    \hline
        1 & $0.71^{+0.06}_{-0.09}$(35) & $0.88^{+0.04}_{-0.13}$(16) & $0.76^{+0.06}_{-0.09}$(29)\\
        2 & $0.30^{+0.09}_{-0.07}$(30) & $0.60^{+0.08}_{-0.09}$(30) & $0.47^{+0.09}_{-0.08}$(32)\\
        3 & $0.18^{+0.07}_{-0.04}$(45) & $0.35^{+0.10}_{-0.08}$(26) & $0.26^{+0.09}_{-0.06}$(35)\\
        4 & $0.19^{+0.09}_{-0.05}$(32) & $0.11^{+0.09}_{-0.04}$(27) & $0.16^{+0.07}_{-0.04}$(43)\\
        5 & $0.00^{+0.10}$(16) & $0.06^{+0.12}_{-0.02}$(16) & $0.19^{+0.13}_{-0.06}$(16)\\
        6 & $0.20^{+0.25}_{-0.08}$(5) & $0.33^{+0.21}_{-0.13}$(6) & $0.00^{+0.21}$(7)\\
        7 & $0.00^{+0.37}$(3) & $0.00^{+0.31}$(4) & -\\
        8 & $0.12^{+0.20}_{-0.04}$(8) & $0.11^{+0.18}_{-0.04}$(9) & $0.17^{+0.23}_{-0.06}$(6)\\
        Out & $0.23^{+0.05}_{-0.04}$(77) & $0.19^{+0.05}_{-0.04}$(74) & $0.19^{+0.06}_{-0.04}$(63)\\
    \hline
         & \multicolumn{3}{c}{$10^{13}< M_\mathrm{200,\:crit}/$M$_\odot<10^{14}$}\\
    \hline
        1 & $0.45^{+0.09}_{-0.08}$(31) & $0.44^{+0.12}_{-0.11}$(18) & $0.31^{+0.08}_{-0.06}$(39)\\
        2 & $0.32^{+0.06}_{-0.05}$(68) & $0.44^{+0.09}_{-0.08}$(32) & $0.22^{+0.08}_{-0.05}$(37)\\
        3 & $0.18^{+0.06}_{-0.04}$(56) & $0.19^{+0.08}_{-0.05}$(37) & $0.20^{+0.06}_{-0.04}$(56)\\
        4 & $0.18^{+0.06}_{-0.04}$(55) & $0.18^{+0.07}_{-0.04}$(49) & $0.12^{+0.06}_{-0.03}$(52)\\
        5 & $0.05^{+0.10}_{-0.02}$(20) & $0.28^{+0.12}_{-0.08}$(18) & $0.15^{0.09}_{-0.04}$(27)\\
        6 & $0.44^{+0.16}_{-0.14}$(9) & $0.09^{+0.16}_{-0.03}$(11) & $0.43^{+0.13}_{-0.11}$(14)\\
        7 & $0.00^{+0.46}$(2) & $0.14^{+0.21}_{-0.05}$(7) & $0.40^{+0.22}_{-0.16}$(5)\\
        8 & $0.31^{+0.15}_{-0.1}$(13) & $0.17^{+0.16}_{-0.06}$(12) & $0.41^{+0.11}_{-0.09}$(22)\\
        Out & $0.17^{+0.05}_{-0.03}$(76) & $0.22^{+0.05}_{-0.04}$(91) & $0.11^{+0.06}_{-0.03}$(54)\\
    \hline
         & \multicolumn{3}{c}{$10^{14}< M_\mathrm{200,\:crit}/$M$_\odot<10^{14.7}$}\\
    \hline
        1 & $0.13^{+0.13}_{-0.05}$(15) & $0.07^{+0.13}_{-0.02}$(15) & $0.0^{+0.17}$(9)\\
        2 & $0.09^{+0.07}_{-0.03}$(34) & $0.28^{+0.12}_{-0.08}$(18) & $0.17^{+0.10}_{-0.05}$(24)\\
        3 & $0.15^{+0.11}_{-0.05}$(20) & $0.14^{+0.09}_{-0.04}$(28) & $0.09^{+0.10}_{-0.03}$(23)\\
        4 & $0.05^{+0.09}_{-0.02}$(21) & $0.24^{+0.09}_{-0.06}$(29) & $0.16^{+0.12}_{-0.05}$(19)\\
        5 & $0.31^{+0.13}_{-0.09}$(16) & $0.29^{+0.14}_{-0.09}$(14) & $0.05^{+0.10}_{-0.02}$(19)\\
        6 & $0.0^{+0.26}$(5) & $0.12^{+0.20}_{-0.04}$(8) & $0.20^{+0.25}_{0.08}$(5)\\
        7 & $0.0^{+0.46}$(2) & $0.0^{+0.46}$(2) &$0.0^{+0.6}$(1) \\
        8 & $0.11^{+0.18}_{-0.04}$(9) & $0.43^{+0.18}_{-0.15}$(7) & $0.17^{+0.23}_{-0.06}$(6) \\
        Out & $0.17^{+0.06}_{-0.04}$(59) & $0.26^{+0.09}_{-0.06}$(34) & $0.0^{+0.37}$(3)\\
    \hline
    \end{tabular}
    \label{tab:feature_fraction_zones}
\end{table}

\begin{figure*}
    \centering
    \includegraphics[width=\linewidth]{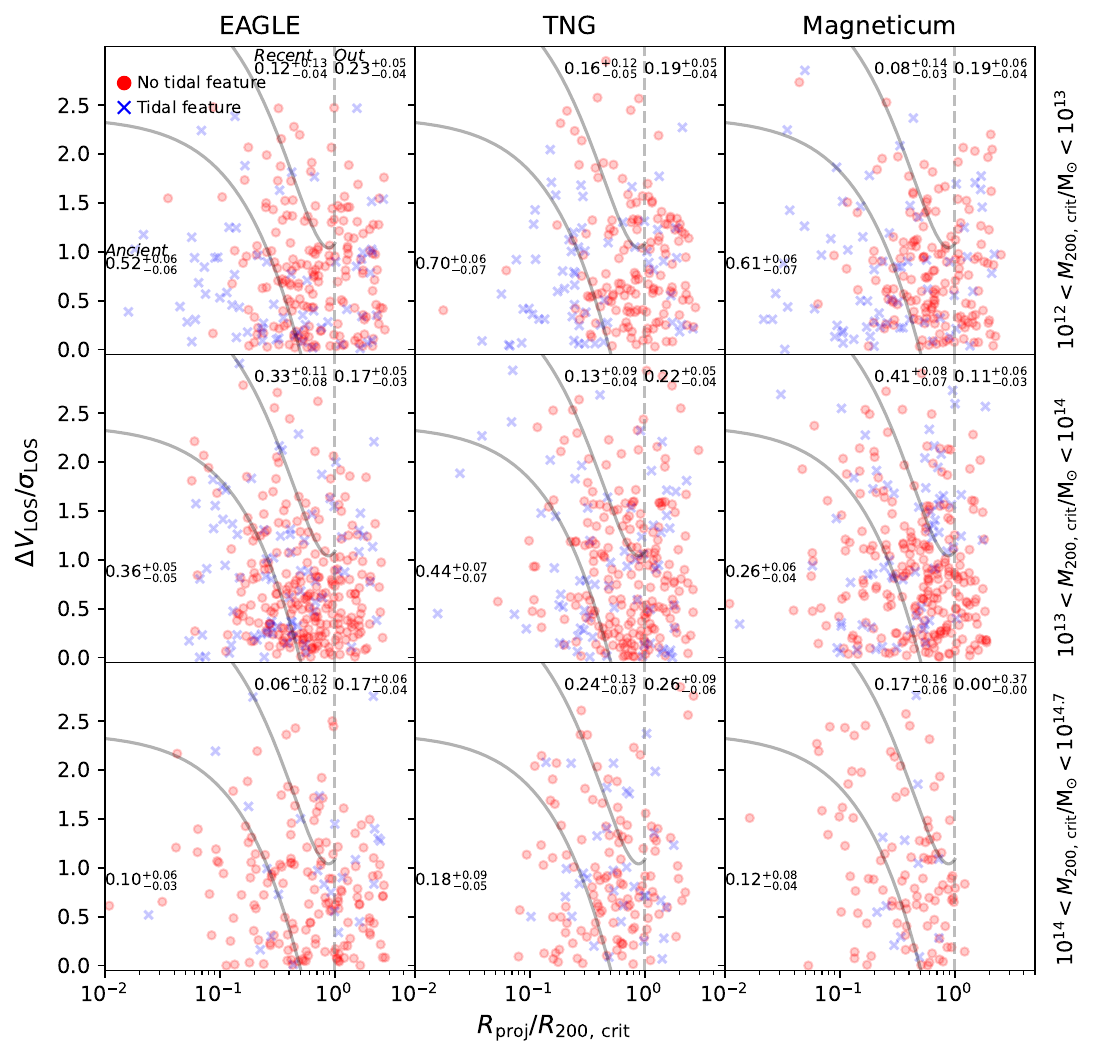}
    \caption{The projected phase-space for satellite galaxies in a range of different halo masses for EAGLE, TNG and Magneticum. The red points represent galaxies without tidal features, and the blue crosses represent galaxies with tidal features. The curves show the zones for ancient infallers ($p\leq2$) and recent infallers ($p\geq5$ and $R_\mathrm{proj}\leq R_\mathrm{200,\:crit}$). The dashed line shows $R=R_{\scriptstyle\mathrm{200,\:crit}}$, the group/cluster outskirts are defined as the region $1\:R_\mathrm{200,\:crit}<R_\mathrm{proj}\leq3\:R_\mathrm{200,\:crt}$. The tidal feature fractions for each zone are provided along with the $1\sigma$ binomial uncertainties. The tidal feature fraction within the ancient infaller zone appears to systematically decrease with increasing halo mass. The tidal feature fraction for recent infallers remains relatively unchanged for all halo masses across all simulations. The ancient infaller fractions for TNG are systematically higher than the ancient infaller fractions for EAGLE and Magneticum for our range of $M_\mathrm{200,\:crit}$.}
    \label{fig:vel_rad_tf}
\end{figure*}

In Fig.~\ref{fig:vel_rad_tf}, we show our sample of satellite galaxies in the projected phase-space across three halo mass bins for each simulation. We provide the tidal feature fractions for each of the zones defined by equation~(\ref{eq:v_r_pasquali}) in Table~\ref{tab:feature_fraction_zones}. Broadly, all simulations show similar trends, with satellites in the recent infaller zone and the outskirts zone beyond $R_{200,\:\mathrm{crit}}$ showing similar tidal feature fractions regardless of the halo mass or the simulation. In contrast, the satellites in the ancient infaller zone tend to start with a higher tidal feature fraction than recent infallers and outskirt satellites in the lowest halo mass bin and decrease with increasing halo mass.

We notice some outliers to the above trend and differences between simulations. The ancient infaller fractions for TNG are systematically (although not always significantly) higher than the ancient infaller fractions for EAGLE and Magneticum for our range of $M_\mathrm{200,\:crit}$. The recent infaller fraction in the $10^{13}<\log_{10}(M_\mathrm{200,\:crit}/\mathrm{M}_\odot)<10^{14}$ bin for TNG is lower than the same fraction in the other simulations. For $10^{14}<\log_{10}(M_\mathrm{200,\:crit}/\mathrm{M}_\odot)<10^{14.7}$, Magneticum lacks tidal feature hosts in the cluster outskirts zones. 

The differences between results across the simulations may reflect differences in the galaxy formation models, resulting in galaxies of different sizes and concentrations, and therefore, different timescales of tidal disruption \citep[e.g.][]{souchayTidesAstronomyAstrophysics2013}. For example, \citet{vandesandeSAMIGalaxySurvey2019} found that while EAGLE has a dearth of galaxies with eccentric/high ellipticities, Magneticum does not. \citet{lagosConnectionMassEnvironment2018} found that the dearth of high ellipticity galaxies was likely due to the implemented cooling floor of $8000\:\mathrm{K}$ in the interstellar medium model, resulting in it being impossible to form a disk thinner than $\sim1\:\mathrm{pkpc}$. A lack of thin disks could result in reduced lifetimes of tidal features \citep[e.g.][]{feldmannTidalDebrisElliptical2008} and lower tidal feature fractions, whereas an excess of thin disks could result in longer tidal feature lifetimes and therefore, higher tidal feature fractions. However, given that we found the occurrence of tidal features as a function of stellar and parent halo mass remains consistent across these three simulations \citep{khalidCharacterizingTidalFeatures2024}, it is likely that small-number statistics are playing a more significant role in driving the variance we see. Considering the incompleteness of our samples, it could be that some of these differences are caused by our sampling. For example, because we sample only a small fraction of the parent halos with $10^{12}<M_\mathrm{200,\:crit}/\mathrm{M}_\odot<10^{13}$ in EAGLE and TNG, this may result in a biased sampling of relaxed (unrelaxed) low mass group halos, where we would expect fewer (more) mergers in the ancient infaller zone. Furthermore, the completeness of our sampling of satellites within each halo drops for $M_\mathrm{200,\:crit}>10^{13}\:\mathrm{M}_\odot$, especially in our TNG sample. The argument that incompleteness is at least partially responsible for the differences we are seeing is supported by the good agreement between EAGLE and Magneticum, which have more similar levels of completeness. Sampling more of the satellite galaxies in each halo and increasing the sample of halos, particularly for $10^{12}<M_\mathrm{200,\:crit}/\mathrm{M}_\odot<10^{13}$, will improve completeness and help test if there are any differences due to subgrid physics models.

To test the statistical significance of the relations in phase-space with halo mass, we require a larger sample size. Therefore, we combine the samples and test whether there are trends in tidal feature fraction in the projected phase-space with halo mass. In doing so, we do lose the potential to probe differences between subgrid physics models. As discussed in Section~\ref{subsec:simulations}, differences in subgrid physics can drive differences in galaxy structure, which could enhance or depress the production of certain tidal features.

% While this does remove our ability to test differences due to different simulation star formation and feedback models, the improved statistics are valuable.

\begin{figure*}
    \centering
    \includegraphics[width=\linewidth]{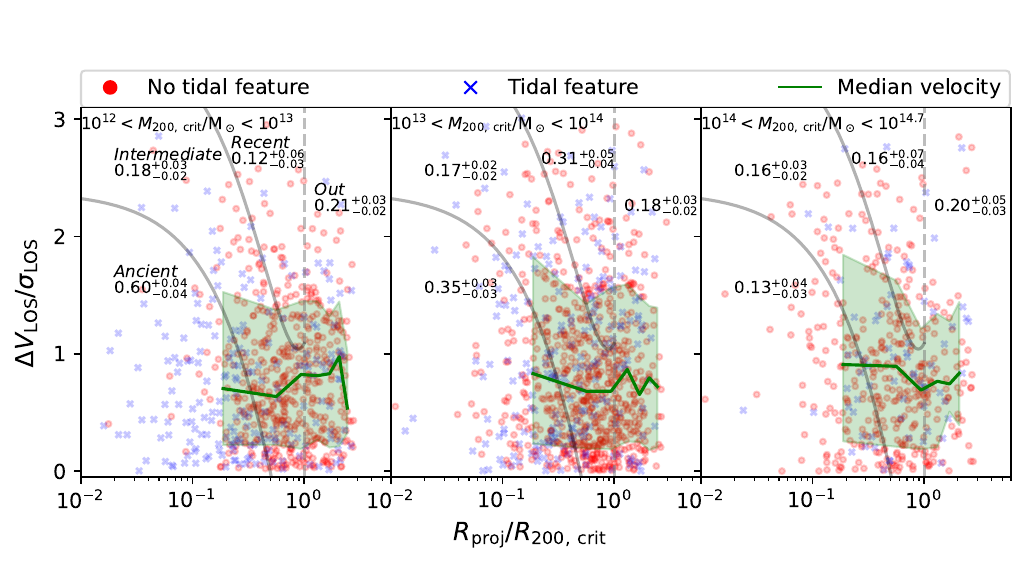}
    \caption{The projected phase-space diagram for the combined sample of satellite galaxies from EAGLE, TNG and Magneticum, across three halo mass bins $\log_{10}(M_{\scriptstyle\mathrm{200,\:crit}}/$M$_{\scriptstyle\odot})=[12,\:13,\:14,\:14.7]$. Galaxies hosting tidal features are plotted with blue crosses, whereas galaxies without tidal features are plotted with red circles. The green line shows the median cluster-centric line-of-sight velocities of the galaxies in 9 radial bins equally spaced between 0 and 1.2 $R_{\scriptstyle\mathrm{200,\:crit}}$, and the shaded green region shows the $16^\mathrm{th}$ and $84^\mathrm{th}$ percentiles. We plot the grey curves corresponding to the ancient infaller and recent infaller zones, the dashed line shows $R_{\mathrm{200,\:crit}}$. There are statistically significant differences between ancient and recent infallers as a function of halo mass. In the lowest halo mass bin, ancient infallers have higher fractions, whereas recent infallers have lower fractions. With increasing halo mass, the ancient infaller fractions decrease while the recent infaller fractions remain the same.}
    \label{fig:vel_rad_tf_combined}
\end{figure*}

In Fig.~\ref{fig:vel_rad_tf_combined}, we show the combined sample of satellite galaxies in the projected phase-space across three halo mass bins. We show the fractions in each halo mass bin for recent and ancient infallers. The fractions for the outskirts galaxies remain consistent at $f_\mathrm{Tidal}\sim0.2$ across the three halo mass bins. Recent infallers and intermediate zone satellites show similar fractions to the outskirts with increasing halo mass, scattering about $f_\mathrm{Tidal}\sim0.2$. The fractions for recent infallers and galaxies in the intermediate zone do scatter to slightly larger values than the outskirts in the $10^{13}<M_\mathrm{200,\:crit}/\mathrm{M}_\odot<10^{14}$ and lower values in $10^{14}<M_\mathrm{200,\:crit}/\mathrm{M}_\odot<10^{14.7}$. However, these differences are not too large, and we find that in unprojected (3D) space the fractions for intermediate zone satellites, recent infallers, and outskirts galaxies become more consistent across all halo masses (Appendix \ref{app:unprojected_phase_space}).

Satellite galaxies in the ancient infaller zone have a significantly higher fraction than recent infallers and outskirt galaxies in the lowest halo mass bin, $f_\mathrm{Tidal}=0.60\pm0.04$. The fraction for satellites in the ancient infaller zone then drops with increasing group/cluster halo mass to be significantly lower $(f_\mathrm{Tidal}=0.13^{+0.04}_{-0.03})$ in the highest halo mass bin. We note that a slight negative gradient in velocity with radius emerges with increasing halo mass. This becomes even clearer in the unprojected results in Appendix \ref{app:unprojected_phase_space}. This suggests that in clusters, the generally higher velocities in the central region may be driving the lack of tidal interactions.

\begin{table}
    \centering
    \caption{The tidal feature fractions with $1\sigma$ binomial confidence intervals and the median halo mass with the 16th and 84th percentiles for the satellites in each halo mass bin and infaller zone and the corresponding sample sizes.\\ $^\mathrm{\ast}$ In the $10^{14}<M_\mathrm{200,\:crit}/\mathrm{M}_\odot<10^{14.7}$ bin, there is only one central.}
    \begin{tabular}{c|c|c}
    \hline
      $\log_{10}(M_\mathrm{200,\:crit}/$M$_\odot)$   & $f_\mathrm{Tidal}$ & Sample size\\
      \hline
        Ancient infallers & & \\
        \hline
        $12.6\pm0.3$ & $0.60\pm0.04$ & 172\\
        $13.48\pm0.4$ & $0.35\pm0.03$ & 225\\
        $14.3\pm0.2$ & $0.13^{+0.04}_{-0.03}$ & 115\\
        \hline
        Recent infallers & &\\
        \hline
        $12.6\pm0.3$ & $0.13^{+0.06}_{-0.03}$ & 48\\
        $13.6^{+0.3}_{-0.4}$ & $0.31\pm^{+0.05}_{-0.04}$ & 95\\
        $14.3\pm0.2$ & $0.16^{+0.07}_{-0.04}$ & 45\\
        \hline
        Outskirts & &\\
        \hline
        $12.6\pm0.3$ & $0.21^{+0.03}_{-0.02}$ & 214\\
        $13.5\pm0.3$ & $0.18^{+0.03}_{-0.02}$ & 221\\
        $14.3^{+0.3}_{-0.1}$ & $0.20^{+0.05}_{-0.03}$ & 96\\
        \hline
        All Satellites & &\\
        \hline
        $12.5\pm0.3$ & $0.29\pm0.02$ & 699\\
        $13.5^{+0.4}_{-0.3}$ & $0.23\pm0.01$ & 927\\
        $14.3\pm0.2$ & $0.16\pm0.02$ & 471\\
        \hline
        Centrals & &\\
        \hline
        $12.3^{+0.3}_{-0.2}$ & $0.48\pm0.02$ & 748\\
        $13.3^{+0.4}_{0.2}$ & $0.76^{+0.04}_{-0.05}$ & 83\\
        $14.6$ & $1_{-0.6}$ & 1$^*$\\
        \hline
        Field galaxies & &\\
        \hline
        $11.5\pm0.3$ & $0.135^{+0.007}_{-0.006}$ & 2864\\
        \hline
    \end{tabular}
    \label{tab:tidal_frac_m200_combined}
\end{table}

\begin{figure}
    \centering
    \includegraphics[width=\linewidth]{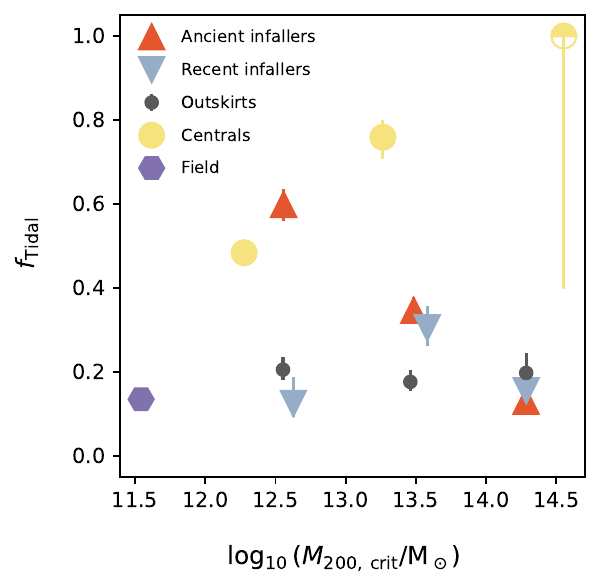}
    \caption{The tidal feature fraction as a function of the group/cluster halo mass for satellite galaxies, separated into the zones for recent (blue downward pointing triangle), ancient (red upward pointing triangle) infallers, centrals (yellow circle), outskirt galaxies with $R_\mathrm{200,\:crit}<R_\mathrm{proj}<3\:R_\mathrm{200,\:crit}$ (black points). We also provide the fraction of galaxies with tidal features for $M_\mathrm{200,\:crit}\leq10^{12}\:\mathrm{M}_\odot$, as an estimate for the field fraction (purple hexagon). The galaxies are binned by halo mass, with bin edges $\log_{10}(M_{\scriptstyle\mathrm{200,\:crit}}/\mathrm{M}_{\scriptstyle\odot})=[12,\:13,\:14,\:14.7]$. We plot the median halo mass and 1 $\sigma$ binomial confidence levels for the tidal feature fractions in these bins. The semi-filled yellow point indicates that the point is based on a single central galaxy. The ancient infallers have a maximum tidal feature fraction at $\log_{10}(M_{\scriptstyle\mathrm{200,\:crit}}/\mathrm{M}_{\scriptstyle\odot})\sim12.5$ and decline beyond this, whereas the recent infallers and outskirts galaxies have fractions that remain roughly consistent with $f_{\scriptstyle\mathrm{Tidal}}\sim0.2\pm0.1$ for all halo masses.}
    \label{fig:tidal_frac_m200_combined}
\end{figure}

\begin{figure}
    \centering
    \includegraphics[width=\linewidth]{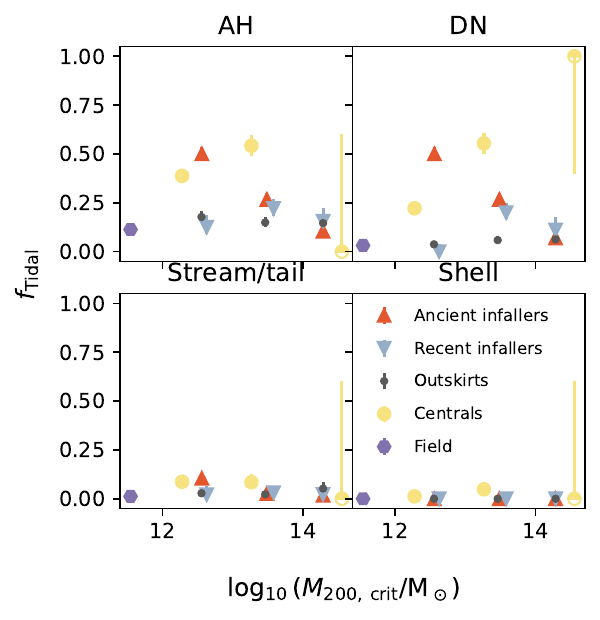}
    \caption{The fraction of galaxies exhibiting each of the tidal feature morphologies as a fraction of the group/cluster halo mass for satellite galaxies, separated into the zones for recent (blue downward pointing triangle), ancient (red upward pointing triangle) infallers, centrals (yellow circle), outskirt galaxies with $1<R_\mathrm{proj}/R_\mathrm{200,\:crit}<3$ (black points). The galaxies are binned by halo mass, with bin edges $\log_{10}(M_{\scriptstyle\mathrm{200,\:crit}}/\mathrm{M}_{\scriptstyle\odot})=[12,\:13,\:14,\:14.7]$. We plot the median halo mass and 1 $\sigma$ binomial confidence intervals for the tidal feature fractions in these bins. The semi-filled yellow point indicates that the point is based on a single central galaxy. The results for stream/tail, asymmetric halo, and double nucleus hosts show qualitatively similar trends to the overall tidal feature fraction behaviour.}
    \label{fig:spec_frac_m200_combined}
\end{figure}

To better characterize the behaviour of tidal feature fractions with halo mass, we summarize the tidal feature fraction as a function of group/cluster mass for satellite galaxies in the ancient infaller zone, recent infaller zone and group/cluster outskirts as a function of halo mass for the combined sample of galaxies across our simulations in Fig.~\ref{fig:tidal_frac_m200_combined}. We provide the measured values in Table~\ref{tab:tidal_frac_m200_combined}. We present the halo mass using the median $M_\mathrm{200,\:crit}$ with the 16th and 84th percentiles. We also measure the tidal feature fraction in haloes with $M_\mathrm{200,\:crit}<10^{12}\mathrm{\:M}_\odot$, as an estimate of the tidal feature fraction for field galaxies.

Fig.~\ref{fig:tidal_frac_m200_combined} shows that satellite galaxies in the group/cluster outskirts have a mean fraction of $f_{\scriptstyle\mathrm{Tidal}}\sim0.2\pm0.1$ across all halo masses. The tidal feature fraction in the recent infaller zone is similar to the outskirts except for the $\log(M_\mathrm{200,\:crit}/\mathrm{M}_\odot)\sim13.6$, where the tidal feature fraction is slightly higher at $0.31^{+0.05}_{-0.04}$. The tidal feature fraction for the outskirts remains higher than the fraction for the field for all halo masses. For recent infallers, the tidal feature fraction remains higher than it is in the field for $10^{13}<M_\mathrm{200,\:crit}/\mathrm{M}_\odot<10^{14.7}$, however, for $10^{12}<M_\mathrm{200,\:crit}/\mathrm{M}_\odot<10^{13}$ it drops to $f_\mathrm{Tidal}=0.13^{+0.04}_{-0.03}$, which is consistent with the field fraction ($f_\mathrm{Tidal}=0.135^{+0.007}_{-0.006}$). However, given the relatively small samples of recent infallers and the unprojected results (Appendix \ref{app:unprojected_phase_space}) showing a higher tidal feature fraction in the recent infaller zone, consistent with the outskirts, we do not find this drop in recent infaller fraction for $10^{14}<M_\mathrm{200,\:crit}/\mathrm{M}_\odot<10^{14.7}$ to be statistically significant. The generally higher tidal feature fractions among the recent infallers and group/cluster outskirt galaxies relative to the field indicate that even the infalling population of satellite galaxies have more tidal feature-producing interactions and mergers than in the field.

Satellite galaxies in the ancient infaller zone start at a fraction of $f_{\scriptstyle\mathrm{Tidal}}=0.60\pm0.04$ at $\log_{10}(M_{\scriptstyle\mathrm{200,\:crit}}/\mathrm{M}_{\scriptstyle\odot})=12.6\pm0.3$, the fraction then decreases with increasing halo mass for halo masses above this down to ${f_{\scriptstyle\mathrm{Tidal}}}=0.13^{+0.04}_{-0.03}$ at $\log_{10}(M_{\scriptstyle\mathrm{200,\:crit}}/\mathrm{M}_{\scriptstyle\odot})=14.3\pm0.2$. The tidal feature fraction for ancient infaller classified satellites is similar to centrals for $M_\mathrm{200,\:crit}\sim10^{12.5}$ M$_\odot$, but at higher halo masses the central galaxies exhibit significantly higher fractions than all satellites. The satellite tidal feature fractions decrease with increasing halo mass for the satellite in the ancient infaller zone, until the fraction is similar to that for recent infallers for cluster-mass haloes ($14<\log_{10}(M_\mathrm{200,\: crit}/$M$_\odot)<14.7$). In our unprojected results (Appendix \ref{app:unprojected_phase_space}), we find that the fractions for ancient infallers into clusters are even lower ($f_\mathrm{Tidal}=0.08^{+0.05}_{-0.02}$) than for recent infallers ($f_\mathrm{Tidal}=0.23^{+0.07}_{-0.05}$) and outskirt satellites ($f_\mathrm{Tidal}=0.18^{+0.04}_{-0.03}$), suggesting a clear suppression of tidal feature occurrence in the cluster centre relative to the outskirts. It is likely that with an even larger sample, this suppression will become clearer in the projected phase space.

Fig.~\ref{fig:spec_frac_m200_combined} shows the fraction of satellite galaxies exhibiting each tidal feature morphology as a function of mean group/cluster halo mass. We have 174 stream/tail host galaxies, 14 shell host galaxies, 1047 asymmetric halo host galaxies and 578 double nucleus galaxies in our total sample, including central and satellite galaxies across all simulations. We find that the frequency of asymmetric halo and double nucleus host galaxies follows similar qualitative trends with halo mass as the overall tidal feature fraction, suggesting that the occurrence of ongoing mergers is driving the trends we see there. Stream/tail hosts do show an increased fraction among satellite galaxies classified as ancient infallers for $M_\mathrm{200,\:crit}\sim10^{12.6}\:\mathrm{M}_\odot$, however, for group and clusters more massive than this, we do not see any differences compared to the stream/tail fraction for recent infallers and outskirt galaxies. Furthermore, the occurrence of stream/tail features around central galaxies remains consistent with increasing halo mass. We also find that shells occur predominantly around central galaxies, with only 1 of the 14 shells occurring around a satellite galaxy. The fraction of central galaxies exhibiting shells is lower than the fraction of central galaxies hosting streams/tails.

In cluster mass halos, we note a slightly lower fraction of double nucleus and asymmetric halo features exhibited by ancient infaller classifed galaxies when compared to those classified as recent infallers. While the difference is not statistically significant here, it becomes so when using the unprojected phase space to classify ancient and recent infallers (Fig.~\ref{fig:spec_frac_m200_unprojected}). The suppression of ancient infaller asymmetric halo host and double nucleus host galaxies relative to recent infaller classified galaxies becomes statistically significant.

Given the incompleteness of our sample, particularly for TNG (Section \ref{subsubsec:completeness}), we further test how completeness impacts our analysis in Appendix \ref{app:completeness} by performing our analysis on a reduced sample. We find that further reducing the completeness of our sample does not qualitatively alter the results presented in this work. This suggests that the nature of the incompleteness in our sample, highlighted in Section \ref{subsubsec:completeness}, is not systematic and therefore does not bias our results.

% Refer back to them as e.g. equation~(\ref{eq:quadratic}).

% Figures are referred to as e.g. Fig.~\ref{fig:example_figure}, and tables as
% e.g. Table~\ref{tab:example_table}.
\section{Discussion}
\label{sec:discussion}

By studying the tidal feature fractions as a function of group/cluster halo mass we found that the peak in the occurrence rate of tidal features at halo masses of $M_{\scriptstyle\mathrm{200,\:crit}}\sim10^{12.5}$ M$_{\scriptstyle\odot}$ is driven primarily by galaxies in the ancient infaller zone as defined by \citet{pasqualiPhysicalPropertiesSDSS2019}. In Section \ref{subsec:tf_mstar}, we investigate if the trends we find could be explained by the relationship between tidal feature fraction and stellar mass found in \citet{khalidCharacterizingTidalFeatures2024}. In Section \ref{subsec:comparison}, we discuss our findings in the context of previous works and explore the possible implications of our results in Section \ref{subsec:implications}.

\subsection{Tidal feature fraction and galaxy stellar mass}
\label{subsec:tf_mstar}

\begin{figure}
    \centering    
    \includegraphics[width=\linewidth]{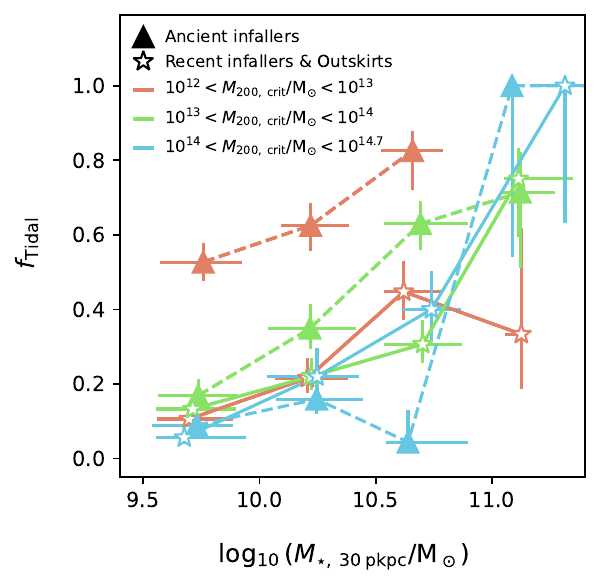}
    \caption{The tidal feature fraction as a function of galaxy stellar mass for satellite galaxies separated into halo mass bins of $\log_{10}(M_{\scriptstyle\mathrm{200,\:crit}}/$M$_{\scriptstyle\odot})=[12,\:13,\:14,\:14.7]$, the results are shown in red, green and blue for the lowest, middle and highest halo mass bins respectively. For improved statistics, we combine the samples for recent infallers and outskirts satellite galaxies. The results for ancient infallers are shown with triangle markers and dashed lines, and the results for recent infallers and outskirts satellite galaxies are shown with star markers and solid lines. We show the points and stellar mass errorbars show the median, 16th and 84th percentile stellar mass, while the tidal feature fraction errorbars show the $1\sigma$ binomial confidence intervals on the tidal feature fraction for each of the 0.5 dex wide stellar mass bins. For satellites in the ancient infallers zone, we find a systematic trend of decreasing tidal feature fraction with increasing halo mass, for all but the $10^{11}<M_{\star,\mathrm{\:30\:pkpc}}/\mathrm{M}_\odot<10^{11.5}$ bin, where the tidal feature fractions appear to converge regardless of halo mass or location in the projected phase-space, albeit with large uncertainties. Recent infallers and outskirts satellite galaxies show no trend with halo mass in the tidal feature fraction stellar mass relationship.}
    \label{fig:tidal_frac_mstar_combined}
\end{figure}

% old caption Tidal feature fraction as a function of stellar mass for satellite galaxies separated into zones for recent (blue points, downard pointing triangle), ancient (red points, upward pointing triangle) infallers, the circular grey points show the relation for all galaxies combined. The galaxies are binned into 11 bins in stellar mass in the range $9.5<\log_{10}(M_{\scriptstyle\star\mathrm{,\:30\:pkpc}}/\mathrm{M}_{\scriptstyle\odot})<12$, with bin widths of 0.25 dex. We plot the median fractions and the median masses in these bins and the $1\sigma$ confidence levels around the medians determined by bootstrapping. The ancient infallers tend to exhibit higher tidal feature fractions than recent infallers for $M_{\scriptstyle\star\mathrm{,\:30\:pkpc}}<10^{10.5}\mathrm{M}_{\scriptstyle\odot}$, whereas recent infallers are consistent with the trend for all satellite galaxies for all stellar masses.

In \citet{khalidCharacterizingTidalFeatures2024}, we found that tidal feature occurrence increased with increasing stellar mass. We also found that the decreasing tidal feature fraction for $M_\mathrm{200,\:crit}\gtrsim10^{12.7}$ M$_\odot$ was driven by satellite galaxies, while the centrals continue to have an increasing tidal feature occurrence with increasing parent halo mass. There was evidence to suggest that the trends seen with halo mass for satellite galaxies could not be explained by the relationship between tidal feature fraction and galaxy stellar mass alone. 

In Fig.~\ref{fig:tidal_frac_mstar_combined}, we disentangle the relationship between tidal feature fraction and stellar mass from the relationship between tidal feature fraction and group/cluster halo mass. We do this to verify that the decreasing tidal feature fraction with increasing halo mass in the ancient infallers and the constant tidal feature fraction with halo mass of recent infallers and group/cluster outskirt galaxies (Fig.~\ref{fig:tidal_frac_m200_combined}) are driven by halo mass. This shows the relationship between tidal feature fraction and stellar mass for recent and ancient infallers, for each of the group/cluster halo mass bins we use in this work ($\log_{10}(M_{\scriptstyle\mathrm{200,\:crit}}/$M$_{\scriptstyle\odot})=[12,\:\:13,\:14,\:14.7]$). We see that for ancient infallers at a fixed stellar mass, there is a strong decrease in tidal feature fraction with increasing group/cluster halo mass for all but the most massive satellites ($M_\mathrm{\star,\:30\:pkpc}>10^{11}\:\mathrm{M}_\odot$) where given our large uncertaintities, we cannot unequivocally state that environment plays a significant role. In contrast, for satellite galaxies in the recent infaller zone, the tidal feature fraction-stellar mass relationship does not change significantly with increasing halo mass.

We have tested that stellar mass is not the driving factor for the differences in tidal feature fraction with Fig.~\ref{fig:vel_rad_mstar} in the Appendix~\ref{app:v_rad_mstar}. We find no significant trends in stellar mass in different zones. If anything, the satellites within the ancient infaller region tend to be slightly more massive, which we would expect to result in an increased tidal feature fraction for our sample \citep{khalidCharacterizingTidalFeatures2024}. As such, our finding of declining tidal feature fraction in the inner regions of groups/clusters with increasing halo mass is not driven by galaxy stellar mass.

Fig.~\ref{fig:tidal_frac_mstar_combined} and Fig.~\ref{fig:vel_rad_mstar} together suggest that the trends we see with ancient infallers and halo mass are not a consequence of the trend with tidal feature fraction and stellar mass, and that the halo mass of the group/cluster plays a significant role. These results suggest that the decreasing tidal feature fraction with halo mass seen for satellite galaxies in \citet{khalidCharacterizingTidalFeatures2024} is driven by the satellite galaxies in the ancient infaller zone.

\subsection{Comparison to previous works}
\label{subsec:comparison}

The decreasing tidal feature fraction with increasing halo mass seen in Fig.~\ref{fig:tidal_frac_m200_combined} for galaxies in the ancient infaller zone is similar to the relationship of the tidal feature fraction as a function of halo mass in \citet{khalidCharacterizingTidalFeatures2024}. In \citet{khalidCharacterizingTidalFeatures2024} the tidal feature fractions peaked at $M_{\scriptstyle\mathrm{200,\:crit}}\sim10^{12.7}$ M$_{\scriptstyle\odot}$, with $f_{\scriptstyle\mathrm{Tidal}}\sim0.4$ for satellite galaxies across the three simulations. In this work, the ancient infallers peak at a similar halo mass of $M_{\scriptstyle\mathrm{200,\:crit}}/\mathrm{M}_\odot=10^{12.6\pm0.3}$, however, they exhibit a much higher mean tidal feature fraction of $f_{\scriptstyle\mathrm{Tidal}}=0.60\pm0.04$, similar to the fraction of centrals hosting tidal features at this halo mass. In contrast, satellites in the recent infaller zone and group/cluster outskirts exhibit a relatively constant fraction with group/cluster halo mass ($f_\mathrm{Tidal}\sim0.2\pm0.1$). Fig.~\ref{fig:spec_frac_m200_combined} shows that the trends seen in Fig.~\ref{fig:tidal_frac_m200_combined} are likely driven by the occurrence of ongoing mergers, as the fraction of asymmetric halo and double nucleus host galaxies reproduce the trends very well.

The suppression of tidal feature fractions in the inner regions of galaxy clusters is also seen in observations. \citet{adamsENVIRONMENTALDEPENDENCEINCIDENCE2012} analysed a sample of 3551 early-type galaxies in 54 low-redshift galaxy clusters and found a suppression of tidal features in the central $\sim0.5\:R_{\scriptstyle\mathrm{200}}$ of the clusters. This trend is qualitatively similar to what we see in Fig.~\ref{fig:vel_rad_tf_combined} for $10^{14}<M_{\scriptstyle\mathrm{200,\:crit}}/\mathrm{M}_{\scriptstyle\odot}<10^{14.7}$. However, \citet{adamsENVIRONMENTALDEPENDENCEINCIDENCE2012} detected far fewer tidal features in general, with just 3\% of the galaxies in their sample exhibiting tidal features. Our higher occurrence rate of tidal features is likely a result of our galaxies all being closer to the mock-observer at $z=0.025$ and our images being deeper by $\sim4$ mag/arcsec$^{2}$ such that fainter tidal features are resolved in our images \citep[e.g.][]{martinPreparingLowSurface2022}. \citet{ohKYDISCGalaxyMorphology2018} also found similar results in their sample of 14 low-redshift Abell clusters, measuring the increasing occurrence of post-merger features (tidal features without any evidence of a companion) with increasing distance from the cluster centre. \citet{sheenPOSTMERGERSIGNATURESREDSEQUENCE2012} and \citet{kimDistributionMergingPostmerger2024} both found that the fraction of ongoing mergers (tidal features indicating an interaction between a galaxy and its companion) was far lower than the fraction of post-merger galaxies within the low-redshift clusters they studied. Consistent with our picture that tidal features from ongoing mergers (double nucleus) are rarer in the cluster environment. \citet{kimDistributionMergingPostmerger2024} also found that ongoing mergers tended to reside within the larger substructures (infalling groups) within the cluster outskirts. This suggests that the tidal features within clusters are formed within infalling groups and in the cluster outskirts, where the relative velocities are low enough for mergers to occur.

The enhancement in the occurrence of tidal interactions in the group environment relative to galaxy clusters was noted in \citet{degerTidalInteractionsMergers2018} using photometric and spectroscopic observations from the ESO Distant Cluster Survey \citep[][]{whiteEDisCSESODistant2005} and \citet{solaLowSurfaceBrightness2025} in deep Canada-France-Hawaii-Telescope data \citep{ferrareseNEXTGENERATIONVIRGO2012,ducATLAS3DProjectXXIX2015,ibataCanadaFranceImaging2017,boselliVirgoEnvironmentalSurvey2018}. In \citet{degerTidalInteractionsMergers2018} sample of 11 galaxy clusters, seven groups and accompanying field galaxies at moderate redshifts, they were able to detect an enhancement of the fraction of galaxies exhibiting merger signatures in groups and cluster outskirts relative to the field and the inner $0.5\:R_\mathrm{200,\:crit}$ of galaxy clusters. They also noted a very slight but not statistically significant increase in interaction rate in the innermost $0.15\:R_\mathrm{200,\:crit}$ of their clusters. We do have an increasing fraction of central galaxies exhibiting tidal features with increasing group/cluster halo mass (Fig.~\ref{fig:tidal_frac_m200_combined}). This could potentially indicate that the increased interaction rate near the cluster centres are driven by interactions with the central galaxies. In their sample of annotated tidal features \citep{solaCharacterizationLowSurface2022}, \citet{solaLowSurfaceBrightness2025} find evidence for enhanced occurrence rates within group-like environments (density of $10^{-1.75}$ galaxies/Mpc$^3$) for all tidal features, and declining rates of occurrence in more dense environments for all tidal features other than tails. This could point to the kinds of mergers that are capable of occurring in clusters. Since tidal tails are likely formed in major mergers \citep[e.g.][]{arpAtlasPeculiarGalaxies1966,toomreGalacticBridgesTails1972}, the results of \citet{solaLowSurfaceBrightness2025} could suggest that only major mergers are capable of occurring in the dense cluster centres, where the relative velocities may be too high for minor mergers to occur. As major mergers are less common than minor mergers, this could be a driving factor in the reduced tidal feature fractions in the centres of our simulated clusters.

The fewer tidal features within the inner regions of the high-mass clusters could be caused by tidal features not contrasting well enough against the intracluster light \citep{martinPreparingLowSurface2022,bazkiaeiDetectionLimitsGalaxy2023,khalidCharacterizingTidalFeatures2024}, which does increase in brightness towards the centre \citep[e.g.][]{montesBuildupIntraclusterLight2021}. The measurement of a decreasing tidal feature fraction with decreasing cluster-centric radius in our deep mock observations and deep observational studies \citep[e.g.][]{sheenPOSTMERGERSIGNATURESREDSEQUENCE2012,ohKYDISCGalaxyMorphology2018} suggests that the trend we find here is real and not a consequence of tidal feature detectability. This supports the argument that we are observing an environmental trend.

Simulations have also found dependence of mergers and satellite survival on cluster-centric distance. \citet{ghignaDarkMatterHaloes1998} studied the properties of dark matter haloes in an N-body simulation of a galaxy cluster and found that between $z=0.5$ and $z=0$, galaxies with apocentres within $\sim0.8\:R_{\scriptstyle\mathrm{200,\:crit}}$ had undergone no mergers within this time, whereas the merger rate outside this radius was $5$ to $9\%$. In the cosmological-hydrodynamical cluster simulation, Hydrangea \citep{baheHydrangeaSimulationsGalaxy2017}, which simulated 24 galaxy clusters with $10^{14}<M^{z=0}_\mathrm{200,\:crit}/\mathrm{M}_\odot<10^{15.4}$, \citet{baheDisruptionSatelliteGalaxies2019} found that satellite galaxy survival to $z=0$ was slightly more common in cluster mass hosts, and crucially 90 per cent of satellite disruption in clusters occurs in lower mass subgroups (pre-processing). These results suggest that the tidal disruption of satellites is significantly more likely to occur in group-mass haloes, and that even the observed disruption of cluster members is most likely due to earlier pre-processing within a galaxy group. \citet{lokasTidalEvolutionGalaxies2020} studied the evolution of galaxies within the most massive cluster in TNG. They found that the mass lost as a result of tidal stripping increased with the number of pericentric passages. This illustrates that in the cluster environment, galaxies that have spent more time on average within the cluster tend to have lost more of their stellar and dark matter mass than galaxies that are more recent infallers. The reduced mass will lead to reduced gravitational force between ancient infallers, which, combined with the high velocities in the centre, would reduce the likelihood of mergers and gravitational interactions extended enough to cause tidal features. 

\subsection{Implications of results}
\label{subsec:implications}

We have found here that the tidal feature fractions of satellite galaxies in the ancient infaller region decrease with increasing halo mass, starting with fractions similar to central galaxies for $10^{12}<M_\mathrm{200,\:crit}/\mathrm{M}_\odot<10^{13}$ and dropping to fractions that are consistent with the recent infallers and outskirts for the cluster mass bin. We show that this trend is mainly driven by the occurrence of double nucleus galaxies and asymmetric halos, suggesting that ongoing mergers decrease with increasing halo mass. We also note that shells occur predominantly around central galaxies, suggesting that the radial mergers that form these features \citep[e.g.][]{amoriscoFeathersBifurcationsShells2015,popFormationIncidenceShell2018,KarademirOuterStellarHalos2019} in groups normally involve a merger with the massive central galaxy. While there is an elevated fraction of stream/tail host galaxies for ancient infallers in groups with $M_\mathrm{200,\:crit}\sim10^{12.6\pm0.6}\:\mathrm{M}_\odot$, they remain consistent with recent infallers and outskirt galaxies for halo masses above this. We also do not see any strong trends with the occurrence of stream/tail features around central galaxies with halo mass. Given the small sample size of stream/tail host galaxies (70), we are unable to draw strong conclusions on the implications of these differences from ongoing mergers on tidal feature survivability post-merger. In Appendix \ref{app:unprojected_phase_space}, we show that in unprojected phase-space, the ancient infallers exhibit even lower fractions than the recent infallers/outskirts galaxies in the highest halo mass bin. From Fig.~\ref{fig:vel_rad_tf_combined} and Fig.~\ref{fig:vel_rad_tf_unprojected}, we see that satellite galaxies in the ancient infaller zone tend to have lower velocities relative to the group/cluster centre in the lowest halo masses and that the velocities in the central region increase with increasing halo mass. Our results suggest that the increasing relative velocities of satellites near the halo centre with increasing halo mass result in a reduced likelihood of mergers and tidal interactions. This leads to our finding of decreasing tidal features with increasing halo mass for ancient infallers. This conclusion is supported by both observations \citep[e.g.][]{adamsENVIRONMENTALDEPENDENCEINCIDENCE2012,sheenPOSTMERGERSIGNATURESREDSEQUENCE2012,ohKYDISCGalaxyMorphology2018,kimDistributionMergingPostmerger2024} and simulations \citep[e.g.][]{ghignaDarkMatterHaloes1998,baheDisruptionSatelliteGalaxies2019,lokasTidalEvolutionGalaxies2020}.  

We also find that the recent infaller/outskirts galaxies show a tidal feature fraction consistently $f_\mathrm{Tidal}\sim0.2\pm0.1$ for all halo masses. Fig.~\ref{fig:tidal_frac_m200_combined} suggests that the peak in the satellite tidal feature fractions seen in \citet{khalidCharacterizingTidalFeatures2024} is driven primarily by the behaviour of galaxies in the ancient infaller zone. Together the results from both the observations \citep{adamsENVIRONMENTALDEPENDENCEINCIDENCE2012,sheenPOSTMERGERSIGNATURESREDSEQUENCE2012,ohKYDISCGalaxyMorphology2018,kimDistributionMergingPostmerger2024} and simulations \citep[][and this work]{ghignaDarkMatterHaloes1998,lokasTidalEvolutionGalaxies2020} suggest the idea that tidal features observed in galaxy clusters are formed from mergers that occur in the lower velocity cluster outskirts, or within groups that eventually fall into the cluster. 

To test if the trends we see with tidal feature fraction and group/cluster halo mass in the recent infaller/outskirt zones can be explained by tidal features brought in by infalling groups, we use a simple toy model. We assume that without infalling groups and clusters, the tidal feature fraction in the recent-infaller and outskirt zones of our groups and clusters will be similar to the tidal feature fraction in the field, $f_\mathrm{Tidal}=0.135^{+0.007}_{-0.006}$ (Table~\ref{tab:feature_fraction_zones}). We estimate the mean richness for our groups and clusters, and present those in Appendix \ref{app:richness}. For a cluster of $M_\mathrm{200,\:crit}\sim10^{14.3}\mathrm{\:M}_\odot$, we would expect a mean richness of $87^{+8}_{-7}$ satellite galaxies with $M_\mathrm{\star,\:30\:pkpc}>10^{9.5}$ M$_\odot$. From Table~\ref{tab:feature_fraction_zones}, we can calculate that 10\% of the satellites are in the recent infaller zone and 20\% reside in the cluster outskirts.

If a group of $M_\mathrm{200,\:crit}\sim10^{13.5}\:\mathrm{M}_\odot$, with a richness of $12.6^{+0.7}_{-0.5}$, was to fall into a cluster of $M_\mathrm{200,\:crit}\sim10^{14.3}\:\mathrm{M}_\odot$, from Table~\ref{tab:tidal_frac_m200_combined}, we estimate it would bring with it $0.23\pm0.01\times12.6^{+0.7}_{-0.5}\simeq2.9\pm0.3$\footnote{When propagating asymmetric uncertainties we take the larger value to provide a conservative estimate of the uncertainty.} tidal feature hosting galaxies. The recent infaller and outskirts zones of the cluster would already have $0.135\times[(0.10+0.20)\times87^{+8}_{-7}]\simeq3.5\pm0.3$ tidal feature hosts. Therefore, as long as none of these features dissipate during the initial infall, the fraction in the recent infaller + outskirts zone of the cluster after the group's infall will be $f_\mathrm{Tidal,\:R+O}\sim0.17\pm0.03$. This is consistent with the fractions we measure for the recent infaller and outskirt zones for clusters.

Similarly, the recent infaller and outskirt zones for groups of mass $10^{13}<M_\mathrm{200,\:crit}<10^{14}\:\mathrm{M}_\odot$, will have 10\% of their satellites in the recent infaller zone and 24\% in the group outskirts. Again, assuming there are no infalling groups, the tidal feature fraction in the recent infaller and outskirt region should be consistent with the field, giving $\sim0.58\pm0.03$ galaxies hosting tidal features. A lower mass group with $M_\mathrm{200,\:crit}\sim10^{12.5}\mathrm{\:M}_\odot$, will have a mean richness of $2.19^{+0.06}_{-0.05}$ galaxies. From Table~\ref{tab:tidal_frac_m200_combined}, we see that this low mass group will have $0.29\pm0.02\times2.19^{+0.06}_{-0.05}\simeq0.64\pm0.06$ tidal features. After the infall of the smaller group into the larger group, this gives us an $f_\mathrm{Tidal,\:R+O}=0.19\pm0.03$, consistent with recent infaller and outskirt fractions of groups of $M_\mathrm{200,\:crit}\sim10^{13.4}\:\mathrm{M}_\odot$. Our toy model shows that infalling groups into higher mass haloes can increase the tidal feature fraction to levels that are consistent with what we measure, suggesting that this is a plausible explanation for our results. 

We note that our toy model is a simplification of the complex dynamical interactions that have taken place between galaxies and their environment over time, which, therefore, might cause predictions to vary from those above. The model does not account for how the lifetime of tidal features depends on the properties of the galaxies involved and the trajectories of the mergers and interactions that created them \citep[e.g.][]{feldmannTidalDebrisElliptical2008,jiLifetimeMergerFeatures2014}. Furthermore, there is no accounting for how the lifetime of tidal features is reduced in the cluster environment, as tidal forces within the cluster potential are effective at stripping post-merger features, reducing their lifetime \citep{jiLifetimeMergerFeatures2014}. The toy model also does not consider how the visibility of the tidal features could be reduced by the contribution of contaminant light (e.g. from nearby galaxies), which could play a role in dense environments such as galaxy clusters \citep[e.g.][]{martinPreparingLowSurface2022,khalidCharacterizingTidalFeatures2024,cuillandreEuclidEarlyRelease2025}.

The decreasing occurrence of tidal features, with decreasing cluster-centric distances and velocities, suggests that these features could be lasting well into their infall. The results of \citet{ghignaDarkMatterHaloes1998,baheDisruptionSatelliteGalaxies2019,lokasTidalEvolutionGalaxies2020} and \citet{kimDistributionMergingPostmerger2024} suggest that the tidal features in our clusters mass haloes ($10^{14}<M_\mathrm{200,\:crit}/\mathrm{M}_\odot<10^{14.7}$) are likely to have formed in mergers in the outskirts or lower halo mass galaxy groups, and the tidal features found within the cluster are likely to be remnants that have remained during infall. If true, this suggests that on average tidal features last $\lesssim3\pm2$ Gyr in the galaxy cluster environment \citep{pasqualiPhysicalPropertiesSDSS2019} as they have mostly dissipated by the time galaxies are ancient infallers. Further investigation of the lifetime of tidal features and how this varies as a function of group/cluster halo mass, halo-centric position and galaxy stellar mass is required to further test this hypothesis. The larger observational samples coming from surveys such as Euclid \citep{euclidcollaborationEuclidOverviewEuclid2025a} and to come from LSST \citep{ivezicLSSTScienceDrivers2019a,robertsonGalaxyFormationEvolution2019,broughVeraRubinObservatory2020} will prove crucial, particularly in constraining the assembly of matter into clusters using the stellar component \citep{kimmigIntraclusterLightDynamical2025a} and disentangling the more subtle potential tensions in the densest environments \citep[e.g.][]{degerTidalInteractionsMergers2018,solaLowSurfaceBrightness2025}.

\section{Conclusions}
\label{sec:conclusions}

We have measured the occurrence rate of visually detected tidal features around satellite galaxies in group and cluster haloes ($12\leq\log_{10}(M_{\scriptstyle\mathrm{200,\:crit}}/\mathrm{M}_{\scriptstyle\odot})\leq14.7$) in the projected phase-space from LSST-like mock images of simulated galaxies. To achieve statistically robust results, we combined the satellite galaxies from three cosmological-hydrodynamical simulations, EAGLE, TNG and Magneticum, where we have previously examined the occurrence rate of tidal features as a function of stellar and halo mass and found that they are consistent across simulations \citep{khalidCharacterizingTidalFeatures2024}. We found that the frequency of tidal features in the ancient infaller zones (defined in \citealt{pasqualiPhysicalPropertiesSDSS2019} as $p\leq2$) has a maximum value of $f_{\scriptstyle\mathrm{Tidal}}=0.60\pm0.04$ at $M_{\scriptstyle\mathrm{200,\:crit}}\sim10^{12.5}$ M$_{\scriptstyle\odot}$ and decreases for halo masses above this, whereas recent infallers ($p\geq5$ and $R\leq R_\mathrm{200,\:crit}$) maintain a roughly constant tidal feature fraction of $f_{\scriptstyle\mathrm{Tidal}}\sim0.2$ for all halo masses, which is consistent with satellites in the group/cluster outskirts beyond ($R_\mathrm{200,\:crit}<R_\mathrm{proj}\leq3\:R_\mathrm{200,\:crit}$). This result shows, for the first time in simulations, that satellite galaxies within the cluster potential show a suppressed tidal feature fraction compared to galaxies outside the cluster or on their initial infall. Thus, replicating what has been noted in observations \citep[e.g.][]{adamsENVIRONMENTALDEPENDENCEINCIDENCE2012,sheenPOSTMERGERSIGNATURESREDSEQUENCE2012,ohKYDISCGalaxyMorphology2018,kimDistributionMergingPostmerger2024}.

These results illustrate that the peak in tidal feature fraction for satellite galaxies, seen in our previous work \citep{khalidCharacterizingTidalFeatures2024}, is driven by the variation in the tidal feature fraction with halo mass for galaxies in the ancient infaller zone. The fact that ancient infallers are likely to have lost a significant portion of their total mass \citep[][]{lokasTidalEvolutionGalaxies2020} and that the velocities at the centres of clusters are higher than they are in groups (Fig.~\ref{fig:vel_rad_tf_combined} and Fig.~\ref{fig:vel_rad_tf_unprojected}) both suggest that the likelihood of tidally disrupting interactions and mergers is reduced within cluster centres. We use a toy model to show that infalling groups can bring a sufficient number of tidal features into a higher halo mass group/cluster to elevate the tidal feature fraction from what we measure for field galaxies ($M_\mathrm{200,\:crit}<10^{12}\:\mathrm{M}_\odot$) to what we measure for group/cluster outskirts and recent infallers. This suggests that the occurrence of tidal features in clusters is primarily driven by pre-processing in galaxy groups. Furthermore, the lack of tidal features in the ancient infaller zone of clusters compared to the recent infaller zones potentially indicates that tidal features that have occurred before or during the initial infall of galaxies into the cluster have largely disappeared by the time the satellite galaxy has become an ancient infaller, putting a limit of $\lesssim3\pm2$ Gyr on the lifetime of such tidal features. Additional work tracking the formation histories of these tidal features could further constrain whether the lifetime of these features varies with halo mass and the impact of that on the observability of tidal features as a function of halo mass. In addition, upcoming LSST observations will enable testing of the predictions made in this work.

\section*{Acknowledgements}

We thank the anonymous reviewer for their thoughtful feedback. We thank Annette Ferguson for helpful feedback. We thank Pierre-Alain Duc for helpful feedback. AK acknowledges the support of an Astronomical Society of Australia Student Travel Grant. SB acknowledges funding support from the Australian Research Council through a Discovery Project DP190101943. RW gratefully acknowledges support from the Forrest Research Foundation. This research includes computations using the computational cluster Katana, supported by Research Technology Services at UNSW Sydney. This work was performed on the OzSTAR national facility at Swinburne University of Technology. The OzSTAR program receives funding in part from the Astronomy National Collaborative Research Infrastructure Strategy (NCRIS) allocation provided by the Australian Government, and from the Victorian Higher Education State Investment Fund (VHESIF) provided by the Victorian Government. We acknowledge the Virgo Consortium for making their simulation data available. The \textsc{eagle} simulations were performed using the DiRAC-2 facility at Durham, managed by the ICC, and the PRACE facility Curie, based in France at TGCC, CEA, Bruyères-le-Châtel. TNG100 was run on the HazelHen Cray XC40 system at the High Performance Computing Center Stuttgart as part of project GCS-ILLU of the Gauss Centre for Supercomputing (GCS). Ancillary and test runs of the IllustrisTNG project were also run on the Stampede supercomputer at TACC/XSEDE (allocation AST140063), at the Hydra and Draco supercomputers at the Max Planck Computing and Data Facility, and on the MIT/Harvard computing facilities supported by FAS and MIT MKI. The {\it Magneticum} simulations were performed at the Leibniz-Rechenzentrum with CPU time assigned to the Project {\it pr83li}. This work was supported by the Deutsche Forschungsgemeinschaft (DFG, German Research Foundation) under Germany's Excellence Strategy - EXC-2094 - 390783311. Parts of this research were supported by the Australian Research Council Centre of Excellence for All Sky Astrophysics in 3 Dimensions (ASTRO 3D), through project number CE170100013.

%%%%%%%%%%%%%%%%%%%%%%%%%%%%%%%%%%%%%%%%%%%%%%%%%%
\section*{Data Availability}

\textsc{EAGLE} data used in this work are publicly available at \url{http://icc.dur.ac.uk/Eagle/}. The \textsc{IllustrisTNG} data used in this work are publicly available at \url{http://www.tng-project.org}. \textsc{Magneticum} data are partially available at \url{https://c2papcosmosim.uc.lrz.de/} \citep{ragagninWebPortalHydrodynamical2017}, with
larger data sets on request.



\bibliographystyle{mnras}
\bibliography{references} % if your bibtex file is called example.bib

\begin{thebibliography}{}
\makeatletter
\relax
\def\mn@urlcharsother{\let\do\@makeother \do\$\do\&\do\#\do\^\do\_\do\%\do\~}
\def\mn@doi{\begingroup\mn@urlcharsother \@ifnextchar [ {\mn@doi@} {\mn@doi@[]}}
\def\mn@doi@[#1]#2{\def\@tempa{#1}\ifx\@tempa\@empty \href {http://dx.doi.org/#2} {doi:#2}\else \href {http://dx.doi.org/#2} {#1}\fi \endgroup}
\def\mn@eprint#1#2{\mn@eprint@#1:#2::\@nil}
\def\mn@eprint@arXiv#1{\href {http://arxiv.org/abs/#1} {{\tt arXiv:#1}}}
\def\mn@eprint@dblp#1{\href {http://dblp.uni-trier.de/rec/bibtex/#1.xml} {dblp:#1}}
\def\mn@eprint@#1:#2:#3:#4\@nil{\def\@tempa {#1}\def\@tempb {#2}\def\@tempc {#3}\ifx \@tempc \@empty \let \@tempc \@tempb \let \@tempb \@tempa \fi \ifx \@tempb \@empty \def\@tempb {arXiv}\fi \@ifundefined {mn@eprint@\@tempb}{\@tempb:\@tempc}{\expandafter \expandafter \csname mn@eprint@\@tempb\endcsname \expandafter{\@tempc}}}

\bibitem[\protect\citeauthoryear{Abadi, Moore  \& Bower}{Abadi et~al.}{1999}]{abadiRamPressureStripping1999}
Abadi M.~G.,  Moore B.,   Bower R.~G.,  1999, \mn@doi [MNRAS] {10.1046/j.1365-8711.1999.02715.x}, 308, 947

\bibitem[\protect\citeauthoryear{Adams, Zaritsky, Sand, Graham, Bildfell, Hoekstra  \& Pritchet}{Adams et~al.}{2012}]{adamsENVIRONMENTALDEPENDENCEINCIDENCE2012}
Adams S.~M.,  Zaritsky D.,  Sand D.~J.,  Graham M.~L.,  Bildfell C.,  Hoekstra H.,   Pritchet C.,  2012, \mn@doi [AJ] {10.1088/0004-6256/144/5/128}, 144, 128

\bibitem[\protect\citeauthoryear{Ade et~al.,}{Ade et~al.}{2014}]{adePlanck2013Results2014}
Ade P. a.~R.,  et~al., 2014, \mn@doi [A\&A] {10.1051/0004-6361/201321591}, 571, A16

\bibitem[\protect\citeauthoryear{Ade et~al.,}{Ade et~al.}{2016}]{adePlanck2015Results2016}
Ade P. a.~R.,  et~al., 2016, \mn@doi [A\&A] {10.1051/0004-6361/201525830}, 594, A13

\bibitem[\protect\citeauthoryear{Agertz, Teyssier  \& Moore}{Agertz et~al.}{2011}]{agertzFormationDiscGalaxies2011}
Agertz O.,  Teyssier R.,   Moore B.,  2011, \mn@doi [MNRAS] {10.1111/j.1365-2966.2010.17530.x}, 410, 1391

\bibitem[\protect\citeauthoryear{Amorisco}{Amorisco}{2015}]{amoriscoFeathersBifurcationsShells2015}
Amorisco N.~C.,  2015, \mn@doi [MNRAS] {10.1093/mnras/stv648}, 450, 575

\bibitem[\protect\citeauthoryear{Arp}{Arp}{1966}]{arpAtlasPeculiarGalaxies1966}
Arp H.,  1966, \mn@doi [ApJS] {10.1086/190147}, 14, 1

\bibitem[\protect\citeauthoryear{Bah{\'e} et~al.,}{Bah{\'e} et~al.}{2017}]{baheHydrangeaSimulationsGalaxy2017}
Bah{\'e} Y.~M.,  et~al., 2017, \mn@doi [MNRAS] {10.1093/mnras/stx1403}, 470, 4186

\bibitem[\protect\citeauthoryear{Bah{\'e} et~al.,}{Bah{\'e} et~al.}{2019}]{baheDisruptionSatelliteGalaxies2019}
Bah{\'e} Y.~M.,  et~al., 2019, \mn@doi [MNRAS] {10.1093/mnras/stz361}, 485, 2287

\bibitem[\protect\citeauthoryear{Barnes}{Barnes}{1988}]{barnesEncountersDiskHalo1988}
Barnes J.~E.,  1988, \mn@doi [ApJ] {10.1086/166593}, 331, 699

\bibitem[\protect\citeauthoryear{Barsanti et~al.,}{Barsanti et~al.}{2018}]{barsantiGalaxyMassAssembly2018}
Barsanti S.,  et~al., 2018, \mn@doi [ApJ] {10.3847/1538-4357/aab61a}, 857, 71

\bibitem[\protect\citeauthoryear{Bazkiaei}{Bazkiaei}{2023}]{bazkiaeiDetectionLimitsGalaxy2023}
Bazkiaei A.~E.,  2023, Thesis, Macquarie University, \mn@doi{10.25949/23974797.v1}

\bibitem[\protect\citeauthoryear{B{\'i}lek et~al.,}{B{\'i}lek et~al.}{2020}]{bilekCensusClassificationLowsurfacebrightness2020}
B{\'i}lek M.,  et~al., 2020, \mn@doi [MNRAS] {10.1093/mnras/staa2248}, 498, 2138

\bibitem[\protect\citeauthoryear{Boselli et~al.,}{Boselli et~al.}{2018}]{boselliVirgoEnvironmentalSurvey2018}
Boselli A.,  et~al., 2018, \mn@doi [A\&A] {10.1051/0004-6361/201732407}, 614, A56

\bibitem[\protect\citeauthoryear{Bottrell et~al.,}{Bottrell et~al.}{2024}]{bottrellIllustrisTNGHSCSSPImage2024}
Bottrell C.,  et~al., 2024, \mn@doi [MNRAS] {10.1093/mnras/stad2971}, 527, 6506

\bibitem[\protect\citeauthoryear{Bournaud, Jog  \& Combes}{Bournaud et~al.}{2005}]{bournaudGalaxyMergersVarious2005}
Bournaud F.,  Jog C.~J.,   Combes F.,  2005, \mn@doi [A\&A] {10.1051/0004-6361:20042036}, 437, 69

\bibitem[\protect\citeauthoryear{Brough et~al.,}{Brough et~al.}{2020}]{broughVeraRubinObservatory2020}
Brough S.,  et~al., 2020, arXiv e-prints, pp arXiv:2001.11067--arXiv:2001.11067

\bibitem[\protect\citeauthoryear{Bruzual \& Charlot}{Bruzual \& Charlot}{2003}]{bruzualStellarPopulationSynthesis2003}
Bruzual G.,  Charlot S.,  2003, \mn@doi [MNRAS] {10.1046/j.1365-8711.2003.06897.x}, 344, 1000

\bibitem[\protect\citeauthoryear{Bryan \& Norman}{Bryan \& Norman}{1998}]{bryanStatisticalPropertiesXRay1998}
Bryan G.~L.,  Norman M.~L.,  1998, \mn@doi [ApJ] {10.1086/305262}, 495, 80

\bibitem[\protect\citeauthoryear{Cameron}{Cameron}{2011}]{cameronEstimationConfidenceIntervals2011}
Cameron E.,  2011, \mn@doi [PASA] {10.1071/AS10046}, 28, 128

\bibitem[\protect\citeauthoryear{Cannarozzo et~al.,}{Cannarozzo et~al.}{2023}]{cannarozzoContributionSituEx2023}
Cannarozzo C.,  et~al., 2023, \mn@doi [MNRAS] {10.1093/mnras/stac3023}, 520, 5651

\bibitem[\protect\citeauthoryear{Chabrier}{Chabrier}{2003}]{chabrierGalacticStellarSubstellar2003}
Chabrier G.,  2003, \mn@doi [PASP] {10.1086/376392}, 115, 763

\bibitem[\protect\citeauthoryear{Crain et~al.,}{Crain et~al.}{2015}]{crainEAGLESimulationsGalaxy2015}
Crain R.~A.,  et~al., 2015, \mn@doi [MNRAS] {10.1093/mnras/stv725}, 450, 1937

\bibitem[\protect\citeauthoryear{Cuillandre et~al.,}{Cuillandre et~al.}{2025}]{cuillandreEuclidEarlyRelease2025}
Cuillandre J.-C.,  et~al., 2025, \mn@doi [A\&A] {10.1051/0004-6361/202450803}, 697, A6

\bibitem[\protect\citeauthoryear{Davison, Norris, Pfeffer, Davies  \& Crain}{Davison et~al.}{2020}]{davisonEAGLEsViewEx2020}
Davison T.~A.,  Norris M.~A.,  Pfeffer J.~L.,  Davies J.~J.,   Crain R.~A.,  2020, \mn@doi [MNRAS] {10.1093/mnras/staa1816}, 497, 81

\bibitem[\protect\citeauthoryear{Deger et~al.,}{Deger et~al.}{2018}]{degerTidalInteractionsMergers2018}
Deger S.,  et~al., 2018, \mn@doi [ApJ] {10.3847/1538-4357/aaeb87}, 869, 6

\bibitem[\protect\citeauthoryear{Desmons, Brough, {Mart{\'i}nez-Lombilla}, De~Propris, Holwerda  \& {L{\'o}pez-S{\'a}nchez}}{Desmons et~al.}{2023}]{desmonsGalaxyMassAssembly2023}
Desmons A.,  Brough S.,  {Mart{\'i}nez-Lombilla} C.,  De~Propris R.,  Holwerda B.,   {L{\'o}pez-S{\'a}nchez} {\'A}.~R.,  2023, \mn@doi [MNRAS] {10.1093/mnras/stad1639}, 523, 4381

\bibitem[\protect\citeauthoryear{Di~Matteo, Springel  \& Hernquist}{Di~Matteo et~al.}{2005}]{dimatteoEnergyInputQuasars2005}
Di~Matteo T.,  Springel V.,   Hernquist L.,  2005, \mn@doi [Nature] {10.1038/nature03335}, 433, 604

\bibitem[\protect\citeauthoryear{Dolag, Borgani, Murante  \& Springel}{Dolag et~al.}{2009}]{dolagSubstructuresHydrodynamicalCluster2009}
Dolag K.,  Borgani S.,  Murante G.,   Springel V.,  2009, \mn@doi [Monthly Notices of the Royal Astronomical Society] {10.1111/j.1365-2966.2009.15034.x}, 399, 497

\bibitem[\protect\citeauthoryear{Dolag et~al.,}{Dolag et~al.}{2025}]{dolagEncyclopediaMagneticumScaling2025}
Dolag K.,  et~al., 2025, Encyclopedia {{Magneticum}}: {{Scaling Relations}} from {{Cosmic Dawn}} to {{Present Day}} (\mn@eprint {arXiv} {2504.01061}), \mn@doi{10.48550/arXiv.2504.01061}

\bibitem[\protect\citeauthoryear{Dubois, Peirani, Pichon, Devriendt, Gavazzi, Welker  \& Volonteri}{Dubois et~al.}{2016}]{duboisHORIZONAGNSimulationMorphological2016}
Dubois Y.,  Peirani S.,  Pichon C.,  Devriendt J.,  Gavazzi R.,  Welker C.,   Volonteri M.,  2016, \mn@doi [MNRAS] {10.1093/mnras/stw2265}, 463, 3948

\bibitem[\protect\citeauthoryear{Dubois et~al.,}{Dubois et~al.}{2021}]{duboisIntroducingNEWHORIZONSimulation2021}
Dubois Y.,  et~al., 2021, \mn@doi [A\&A] {10.1051/0004-6361/202039429}, 651, A109

\bibitem[\protect\citeauthoryear{Duc et~al.,}{Duc et~al.}{2015}]{ducATLAS3DProjectXXIX2015}
Duc P.-A.,  et~al., 2015, \mn@doi [MNRAS] {10.1093/mnras/stu2019}, 446, 120

\bibitem[\protect\citeauthoryear{{Euclid Collaboration} et~al.,}{{Euclid Collaboration} et~al.}{2025}]{euclidcollaborationEuclidOverviewEuclid2025a}
{Euclid Collaboration} et~al., 2025, \mn@doi [A\&A] {10.1051/0004-6361/202450810}, 697, A1

\bibitem[\protect\citeauthoryear{Fall \& Efstathiou}{Fall \& Efstathiou}{1980}]{fallFormationRotationDisc1980}
Fall S.,  Efstathiou G.,  1980, \mn@doi [MNRAS] {10.1093/mnras/193.2.189}, 193, 189

\bibitem[\protect\citeauthoryear{Feldmann, Mayer  \& Carollo}{Feldmann et~al.}{2008}]{feldmannTidalDebrisElliptical2008}
Feldmann R.,  Mayer L.,   Carollo C.~M.,  2008, \mn@doi [ApJ] {10.1086/590235}, 684, 1062

\bibitem[\protect\citeauthoryear{Ferrarese et~al.,}{Ferrarese et~al.}{2012}]{ferrareseNEXTGENERATIONVIRGO2012}
Ferrarese L.,  et~al., 2012, \mn@doi [ApJS] {10.1088/0067-0049/200/1/4}, 200, 4

\bibitem[\protect\citeauthoryear{Ghigna, Moore, Governato, Lake, Quinn  \& Stadel}{Ghigna et~al.}{1998}]{ghignaDarkMatterHaloes1998}
Ghigna S.,  Moore B.,  Governato F.,  Lake G.,  Quinn T.,   Stadel J.,  1998, \mn@doi [MNRAS] {10.1046/j.1365-8711.1998.01918.x}, 300, 146

\bibitem[\protect\citeauthoryear{Gnedin}{Gnedin}{2003}]{gnedinTidalEffectsClusters2003}
Gnedin O.~Y.,  2003, \mn@doi [ApJ] {10.1086/344636}, 582, 141

\bibitem[\protect\citeauthoryear{Gunn \& Gott}{Gunn \& Gott}{1972}]{gunnInfallMatterClusters1972}
Gunn J.~E.,  Gott III J.~R.,  1972, \mn@doi [ApJ] {10.1086/151605}, 176, 1

\bibitem[\protect\citeauthoryear{Hendel \& Johnston}{Hendel \& Johnston}{2015}]{hendelTidalDebrisMorphology2015}
Hendel D.,  Johnston K.~V.,  2015, \mn@doi [MNRAS] {10.1093/mnras/stv2035}, 454, 2472

\bibitem[\protect\citeauthoryear{Hernquist \& Barnes}{Hernquist \& Barnes}{1991}]{hernquistOriginKinematicSubsystems1991}
Hernquist L.,  Barnes J.~E.,  1991, \mn@doi [Nature] {10.1038/354210a0}, 354, 210

\bibitem[\protect\citeauthoryear{Hopkins et~al.,}{Hopkins et~al.}{2010a}]{hopkinsMergersBulgeFormation2010a}
Hopkins P.~F.,  et~al., 2010a, \mn@doi [ApJ] {10.1088/0004-637X/715/1/202}, 715, 202

\bibitem[\protect\citeauthoryear{Hopkins et~al.,}{Hopkins et~al.}{2010b}]{hopkinsMERGERSLCDMUNCERTAINTIES2010}
Hopkins P.~F.,  et~al., 2010b, \mn@doi [ApJ] {10.1088/0004-637X/724/2/915}, 724, 915

\bibitem[\protect\citeauthoryear{Huang \& Fan}{Huang \& Fan}{2022}]{huangMassiveEarlyTypeGalaxies2022}
Huang Q.,  Fan L.,  2022, \mn@doi [ApJS] {10.3847/1538-4365/ac85b1}, 262, 39

\bibitem[\protect\citeauthoryear{Ibata et~al.,}{Ibata et~al.}{2017}]{ibataCanadaFranceImaging2017}
Ibata R.~A.,  et~al., 2017, \mn@doi [ApJ] {10.3847/1538-4357/aa855c}, 848, 128

\bibitem[\protect\citeauthoryear{Ivezi{\'c} et~al.,}{Ivezi{\'c} et~al.}{2019}]{ivezicLSSTScienceDrivers2019a}
Ivezi{\'c} {\v Z}.,  et~al., 2019, \mn@doi [ApJ] {10.3847/1538-4357/ab042c}, 873, 111

\bibitem[\protect\citeauthoryear{Ji, Peirani  \& Yi}{Ji et~al.}{2014}]{jiLifetimeMergerFeatures2014}
Ji I.,  Peirani S.,   Yi S.~K.,  2014, \mn@doi [A\&A] {10.1051/0004-6361/201423530}, 566, A97

\bibitem[\protect\citeauthoryear{Jian, Lin  \& Chiueh}{Jian et~al.}{2012}]{jianEnvironmentalDependenceGalaxy2012}
Jian H.-Y.,  Lin L.,   Chiueh T.,  2012, \mn@doi [ApJ] {10.1088/0004-637X/754/1/26}, 754, 26

\bibitem[\protect\citeauthoryear{Karademir, Remus, Burkert, Dolag, Hoffmann, Moster, Steinwandel  \& Zhang}{Karademir et~al.}{2019}]{KarademirOuterStellarHalos2019}
Karademir G.~S.,  Remus R.-S.,  Burkert A.,  Dolag K.,  Hoffmann T.~L.,  Moster B.~P.,  Steinwandel U.~P.,   Zhang J.,  2019, \mn@doi [MNRAS] {10.1093/mnras/stz1251}, 487, 318

\bibitem[\protect\citeauthoryear{Khalid, Brough, Martin, Kimmig, Lagos, Remus  \& {Martinez-Lombilla}}{Khalid et~al.}{2024}]{khalidCharacterizingTidalFeatures2024}
Khalid A.,  Brough S.,  Martin G.,  Kimmig L.~C.,  Lagos C. D.~P.,  Remus R.~S.,   {Martinez-Lombilla} C.,  2024, \mn@doi [MNRAS] {10.1093/mnras/stae1064}, 530, 4422

\bibitem[\protect\citeauthoryear{Kim et~al.,}{Kim et~al.}{2024}]{kimDistributionMergingPostmerger2024}
Kim D.,  et~al., 2024, \mn@doi [ApJ] {10.3847/1538-4357/ad32ce}, 966, 124

\bibitem[\protect\citeauthoryear{Kimmig et~al.,}{Kimmig et~al.}{2025}]{kimmigIntraclusterLightDynamical2025a}
Kimmig L.~C.,  et~al., 2025, \mn@doi [A\&A] {10.1051/0004-6361/202554777}, 700, A95

\bibitem[\protect\citeauthoryear{Komatsu et~al.,}{Komatsu et~al.}{2011}]{komatsuSevenyearWilkinsonMicrowave2011}
Komatsu E.,  et~al., 2011, \mn@doi [ApJS] {10.1088/0067-0049/192/2/18}, 192, 18

\bibitem[\protect\citeauthoryear{Lagos et~al.,}{Lagos et~al.}{2018a}]{lagosQuantifyingImpactMergers2018}
Lagos C. d.~P.,  et~al., 2018a, \mn@doi [MNRAS] {10.1093/mnras/stx2667}, 473, 4956

\bibitem[\protect\citeauthoryear{Lagos, Schaye, Bah{\'e}, {Van de Sande}, Kay, Barnes, Davis  \& Dalla~Vecchia}{Lagos et~al.}{2018b}]{lagosConnectionMassEnvironment2018}
Lagos C. d.~P.,  Schaye J.,  Bah{\'e} Y.,  {Van de Sande} J.,  Kay S.~T.,  Barnes D.,  Davis T.~A.,   Dalla~Vecchia C.,  2018b, \mn@doi [MNRAS] {10.1093/mnras/sty489}, 476, 4327

\bibitem[\protect\citeauthoryear{Lewis et~al.,}{Lewis et~al.}{2002}]{lewis2dFGalaxyRedshift2002}
Lewis I.,  et~al., 2002, \mn@doi [MNRAS] {10.1046/j.1365-8711.2002.05558.x}, 334, 673

\bibitem[\protect\citeauthoryear{{\L}okas}{{\L}okas}{2020}]{lokasTidalEvolutionGalaxies2020}
{\L}okas E.~L.,  2020, \mn@doi [A\&A] {10.1051/0004-6361/202037643}, 638, A133

\bibitem[\protect\citeauthoryear{Ludlow, Fall, Schaye  \& Obreschkow}{Ludlow et~al.}{2021}]{ludlowSpuriousHeatingStellar2021}
Ludlow A.~D.,  Fall S.~M.,  Schaye J.,   Obreschkow D.,  2021, \mn@doi [MNRAS] {10.1093/mnras/stab2770}, 508, 5114

\bibitem[\protect\citeauthoryear{Ludlow, Fall, Wilkinson, Schaye  \& Obreschkow}{Ludlow et~al.}{2023}]{ludlowSpuriousHeatingStellar2023}
Ludlow A.~D.,  Fall S.~M.,  Wilkinson M.~J.,  Schaye J.,   Obreschkow D.,  2023, \mn@doi [MNRAS] {10.1093/mnras/stad2615}, 525, 5614

\bibitem[\protect\citeauthoryear{Malin \& Carter}{Malin \& Carter}{1983}]{malinCatalogEllipticalGalaxies1983}
Malin D.~F.,  Carter D.,  1983, \mn@doi [ApJ] {10.1086/161467}, 274, 534

\bibitem[\protect\citeauthoryear{Mancillas, Duc, Combes, Bournaud, Emsellem, Martig  \& {Michel-Dansac}}{Mancillas et~al.}{2019}]{mancillasProbingMergerHistory2019}
Mancillas B.,  Duc P.-A.,  Combes F.,  Bournaud F.,  Emsellem E.,  Martig M.,   {Michel-Dansac} L.,  2019, \mn@doi [A\&A] {10.1051/0004-6361/201936320}, 632, A122

\bibitem[\protect\citeauthoryear{Marinacci et~al.,}{Marinacci et~al.}{2018}]{marinacciFirstResultsIllustrisTNG2018}
Marinacci F.,  et~al., 2018, \mn@doi [MNRAS] {10.1093/mnras/sty2206}, 480, 5113

\bibitem[\protect\citeauthoryear{Martin, Kaviraj, Devriendt, Dubois, Laigle  \& Pichon}{Martin et~al.}{2017}]{martinLimitedRoleGalaxy2017}
Martin G.,  Kaviraj S.,  Devriendt J. E.~G.,  Dubois Y.,  Laigle C.,   Pichon C.,  2017, \mn@doi [MNRAS] {10.1093/mnrasl/slx136}, 472, L50

\bibitem[\protect\citeauthoryear{Martin, Kaviraj, Devriendt, Dubois  \& Pichon}{Martin et~al.}{2018}]{martinRoleMergersDriving2018}
Martin G.,  Kaviraj S.,  Devriendt J. E.~G.,  Dubois Y.,   Pichon C.,  2018, \mn@doi [MNRAS] {10.1093/mnras/sty1936}, 480, 2266

\bibitem[\protect\citeauthoryear{Martin et~al.,}{Martin et~al.}{2021}]{martinRoleMergersInteractions2021}
Martin G.,  et~al., 2021, \mn@doi [MNRAS] {10.1093/mnras/staa3443}, 500, 4937

\bibitem[\protect\citeauthoryear{Martin et~al.,}{Martin et~al.}{2022}]{martinPreparingLowSurface2022}
Martin G.,  et~al., 2022, \mn@doi [MNRAS] {10.1093/mnras/stac1003}, 513, 1459

\bibitem[\protect\citeauthoryear{Martin, Pearce, Hatch, {Contreras-Santos}, Knebe  \& Cui}{Martin et~al.}{2024}]{martinStellarStrippingEfficiencies2024}
Martin G.,  Pearce F.~R.,  Hatch N.~A.,  {Contreras-Santos} A.,  Knebe A.,   Cui W.,  2024, \mn@doi [MNRAS] {10.1093/mnras/stae2488}, 535, 2375

\bibitem[\protect\citeauthoryear{Merritt et~al.,}{Merritt et~al.}{2020}]{merrittMissingOutskirtsProblem2020}
Merritt A.,  et~al., 2020, \mn@doi [MNRAS] {10.1093/MNRAS/STAA1164}, 495, 4570

\bibitem[\protect\citeauthoryear{Mihos}{Mihos}{2003}]{mihosInteractionsMergersCluster2003}
Mihos C.,  2003, \mn@doi [arXiv e-prints] {10.48550/arXiv.astro-ph/0305512}, pp astro--ph/0305512

\bibitem[\protect\citeauthoryear{Montes, Brough, Owers  \& Santucci}{Montes et~al.}{2021}]{montesBuildupIntraclusterLight2021}
Montes M.,  Brough S.,  Owers M.~S.,   Santucci G.,  2021, \mn@doi [ApJ] {10.3847/1538-4357/abddb6}, 910, 45

\bibitem[\protect\citeauthoryear{Naab, Johansson  \& Ostriker}{Naab et~al.}{2009}]{naabMINORMERGERSSIZE2009}
Naab T.,  Johansson P.~H.,   Ostriker J.~P.,  2009, \mn@doi [ApJ] {10.1088/0004-637X/699/2/L178}, 699, L178

\bibitem[\protect\citeauthoryear{Naiman et~al.,}{Naiman et~al.}{2018}]{naimanFirstResultsIllustrisTNG2018}
Naiman J.~P.,  et~al., 2018, \mn@doi [MNRAS] {10.1093/mnras/sty618}, 477, 1206

\bibitem[\protect\citeauthoryear{Nelson et~al.,}{Nelson et~al.}{2018}]{nelsonFirstResultsIllustrisTNG2018}
Nelson D.,  et~al., 2018, \mn@doi [MNRAS] {10.1093/mnras/stx3040}, 475, 624

\bibitem[\protect\citeauthoryear{Nipoti}{Nipoti}{2025}]{nipotiEvolutionMassiveQuiescent2025}
Nipoti C.,  2025, Evolution of Massive Quiescent Galaxies via Envelope Accretion (\mn@eprint {arXiv} {2502.19497}), \mn@doi{10.48550/arXiv.2502.19497}

\bibitem[\protect\citeauthoryear{Oh et~al.,}{Oh et~al.}{2018}]{ohKYDISCGalaxyMorphology2018}
Oh S.,  et~al., 2018, \mn@doi [ApJS] {10.3847/1538-4365/aacd47}, 237, 14

\bibitem[\protect\citeauthoryear{Olivier, Seppala  \& Gilmore}{Olivier et~al.}{2008}]{olivierOpticalDesignLSST2008a}
Olivier S.~S.,  Seppala L.,   Gilmore K.,  2008, in {Atad-Ettedgui} E.,  Lemke D.,  eds, {{SPIE Astronomical Telescopes}} + {{Instrumentation}}. Marseille, France, p. 70182G, \mn@doi{10.1117/12.790264}

\bibitem[\protect\citeauthoryear{Omori et~al.,}{Omori et~al.}{2023}]{omoriGalaxyMergersSubaru2023b}
Omori K.~C.,  et~al., 2023, \mn@doi [A\&A] {10.1051/0004-6361/202346743}, 679, A142

\bibitem[\protect\citeauthoryear{Ostriker \& Cowie}{Ostriker \& Cowie}{1981}]{ostrikerGalaxyFormationIntergalactic1981}
Ostriker J.~P.,  Cowie L.~L.,  1981, \mn@doi [ApJ] {10.1086/183458}, 243, L127

\bibitem[\protect\citeauthoryear{Pasquali, Smith, Gallazzi, De~Lucia, Zibetti, Hirschmann  \& Yi}{Pasquali et~al.}{2019}]{pasqualiPhysicalPropertiesSDSS2019}
Pasquali A.,  Smith R.,  Gallazzi A.,  De~Lucia G.,  Zibetti S.,  Hirschmann M.,   Yi S.~K.,  2019, \mn@doi [MNRAS] {10.1093/mnras/sty3530}, 484, 1702

\bibitem[\protect\citeauthoryear{Pe{\~n}arrubia, Benson, Walker, Gilmore, McConnachie  \& Mayer}{Pe{\~n}arrubia et~al.}{2010}]{penarrubiaImpactDarkMatter2010}
Pe{\~n}arrubia J.,  Benson A.~J.,  Walker M.~G.,  Gilmore G.,  McConnachie A.~W.,   Mayer L.,  2010, \mn@doi [MNRAS] {10.1111/j.1365-2966.2010.16762.x}, 406, 1290

\bibitem[\protect\citeauthoryear{Petrosian}{Petrosian}{1976}]{petrosianSurfaceBrightnessEvolution1976}
Petrosian V.,  1976, \mn@doi [ApJ] {10.1086/182301}, 210, L53

\bibitem[\protect\citeauthoryear{Pillepich et~al.,}{Pillepich et~al.}{2018a}]{pillepichSimulatingGalaxyFormation2018}
Pillepich A.,  et~al., 2018a, \mn@doi [MNRAS] {10.1093/mnras/stx2656}, 473, 4077

\bibitem[\protect\citeauthoryear{Pillepich et~al.,}{Pillepich et~al.}{2018b}]{pillepichFirstResultsIllustristng2018}
Pillepich A.,  et~al., 2018b, \mn@doi [MNRAS] {10.1093/mnras/stx3112}, 475, 648

\bibitem[\protect\citeauthoryear{Pop, Pillepich, Amorisco  \& Hernquist}{Pop et~al.}{2018}]{popFormationIncidenceShell2018}
Pop A.-R.,  Pillepich A.,  Amorisco N.~C.,   Hernquist L.,  2018, \mn@doi [MNRAS] {10.1093/mnras/sty1932}, 480, 1715

\bibitem[\protect\citeauthoryear{Popesso et~al.,}{Popesso et~al.}{2024}]{popessoAverageXrayProperties2024}
Popesso P.,  et~al., 2024, Average {{X-ray}} Properties of Galaxy Groups. {{From Milky Way-like}} Halos to Massive Clusters, \mn@doi{10.48550/arXiv.2411.17120}

\bibitem[\protect\citeauthoryear{Press \& Schechter}{Press \& Schechter}{1974}]{pressFormationGalaxiesClusters1974}
Press W.~H.,  Schechter P.,  1974, \mn@doi [ApJ] {10.1086/152650}, 187, 425

\bibitem[\protect\citeauthoryear{Ragagnin, Dolag, Biffi, Cadolle~Bel, Hammer, Krukau, Petkova  \& Steinborn}{Ragagnin et~al.}{2017}]{ragagninWebPortalHydrodynamical2017}
Ragagnin A.,  Dolag K.,  Biffi V.,  Cadolle~Bel M.,  Hammer N.~J.,  Krukau A.,  Petkova M.,   Steinborn D.,  2017, \mn@doi [Astronomy and Computing, Volume 20, p. 52-67.] {10.1016/j.ascom.2017.05.001}, 20, 52

\bibitem[\protect\citeauthoryear{Remus \& Forbes}{Remus \& Forbes}{2022}]{remusAccretedNotAccreted2022}
Remus R.-S.,  Forbes D.~A.,  2022, \mn@doi [ApJ] {10.3847/1538-4357/ac7b30}, 935, 37

\bibitem[\protect\citeauthoryear{Rhee, Smith, Choi, Yi, Jaff{\'e}, Candlish  \& {S{\'a}nchez-J{\'a}nssen}}{Rhee et~al.}{2017}]{rheePhasespaceAnalysisGroup2017}
Rhee J.,  Smith R.,  Choi H.,  Yi S.~K.,  Jaff{\'e} Y.,  Candlish G.,   {S{\'a}nchez-J{\'a}nssen} R.,  2017, \mn@doi [ApJ] {10.3847/1538-4357/aa6d6c}, 843, 128

\bibitem[\protect\citeauthoryear{Robertson et~al.,}{Robertson et~al.}{2019}]{robertsonGalaxyFormationEvolution2019}
Robertson B.~E.,  et~al., 2019, \mn@doi [Nature Reviews Physics] {10.1038/s42254-019-0067-x}, 1, 450

\bibitem[\protect\citeauthoryear{{Rodriguez-Gomez} et~al.,}{{Rodriguez-Gomez} et~al.}{2015}]{rodriguez-gomezMergerRateGalaxies2015}
{Rodriguez-Gomez} V.,  et~al., 2015, \mn@doi [MNRAS] {10.1093/mnras/stv264}, 449, 49

\bibitem[\protect\citeauthoryear{Rom{\'a}n, Trujillo  \& Montes}{Rom{\'a}n et~al.}{2020}]{romanGalacticCirriDeep2020}
Rom{\'a}n J.,  Trujillo I.,   Montes M.,  2020, \mn@doi [A\&A] {10.1051/0004-6361/201936111}, 644, A42

\bibitem[\protect\citeauthoryear{Rutherford et~al.,}{Rutherford et~al.}{2024}]{rutherfordSAMIGalaxySurvey2024}
Rutherford T.~H.,  et~al., 2024, \mn@doi [MNRAS] {10.1093/mnras/stae398}, 529, 810

\bibitem[\protect\citeauthoryear{Ryden \& Gunn}{Ryden \& Gunn}{1987}]{rydenGALAXYFORMATIONGRAVITATIONAL1987}
Ryden B.~S.,  Gunn J.~E.,  1987, ApJ, 318, 15

\bibitem[\protect\citeauthoryear{Schaye et~al.,}{Schaye et~al.}{2015}]{schayeEAGLEProjectSimulating2015}
Schaye J.,  et~al., 2015, \mn@doi [MNRAS] {10.1093/mnras/stu2058}, 446, 521

\bibitem[\protect\citeauthoryear{Sheen, Yi, Ree  \& Lee}{Sheen et~al.}{2012}]{sheenPOSTMERGERSIGNATURESREDSEQUENCE2012}
Sheen Y.-K.,  Yi S.~K.,  Ree C.~H.,   Lee J.,  2012, \mn@doi [ApJS] {10.1088/0067-0049/202/1/8}, 202, 8

\bibitem[\protect\citeauthoryear{Sola et~al.,}{Sola et~al.}{2022}]{solaCharacterizationLowSurface2022}
Sola E.,  et~al., 2022, \mn@doi [A\&A] {10.1051/0004-6361/202142675}, 662, A124

\bibitem[\protect\citeauthoryear{Sola et~al.,}{Sola et~al.}{2025}]{solaLowSurfaceBrightness2025}
Sola E.,  et~al., 2025, Low {{Surface Brightness}} Structures from Annotated Deep {{CFHT}} Images: Effects of the Host Galaxy's Properties and Environment (\mn@eprint {arXiv} {2503.18480}), \mn@doi{10.48550/arXiv.2503.18480}

\bibitem[\protect\citeauthoryear{Souchay, Mathis  \& Tokieda}{Souchay et~al.}{2013}]{souchayTidesAstronomyAstrophysics2013}
Souchay J.,  Mathis S.,   Tokieda T.,  2013, Tides in {{Astronomy}} and {{Astrophysics}}.
 Lecture {{Notes}} in {{Physics}} Vol. 861, Springer Berlin, Heidelberg, \mn@doi{10.1007/978-3-642-32961-6}

\bibitem[\protect\citeauthoryear{Springel, White, Tormen  \& Kauffmann}{Springel et~al.}{2001}]{springelPopulatingClusterGalaxies2001}
Springel V.,  White S. D.~M.,  Tormen G.,   Kauffmann G.,  2001, \mn@doi [MNRAS] {10.1046/j.1365-8711.2001.04912.x}, 328, 726

\bibitem[\protect\citeauthoryear{Springel, Di~Matteo  \& Hernquist}{Springel et~al.}{2005a}]{springelModellingFeedbackStars2005}
Springel V.,  Di~Matteo T.,   Hernquist L.,  2005a, \mn@doi [MNRAS] {10.1111/j.1365-2966.2005.09238.x}, 361, 776

\bibitem[\protect\citeauthoryear{Springel et~al.,}{Springel et~al.}{2005b}]{springelSimulationsFormationEvolution2005}
Springel V.,  et~al., 2005b, \mn@doi [Nature] {10.1038/nature03597}, 435, 629

\bibitem[\protect\citeauthoryear{Springel et~al.,}{Springel et~al.}{2018}]{springelFirstResultsIllustrisTNG2018}
Springel V.,  et~al., 2018, \mn@doi [MNRAS] {10.1093/mnras/stx3304}, 475, 676

\bibitem[\protect\citeauthoryear{Teklu, Remus, Dolag, Beck, Burkert, Schmidt, Schulze  \& Steinborn}{Teklu et~al.}{2015}]{tekluConnectingAngularMomentum2015}
Teklu A.~F.,  Remus R.-S.,  Dolag K.,  Beck A.~M.,  Burkert A.,  Schmidt A.~S.,  Schulze F.,   Steinborn L.~K.,  2015, \mn@doi [ApJ] {10.1088/0004-637X/812/1/29}, 812, 29

\bibitem[\protect\citeauthoryear{Toomre \& Toomre}{Toomre \& Toomre}{1972}]{toomreGalacticBridgesTails1972}
Toomre A.,  Toomre J.,  1972, \mn@doi [ApJ] {10.1086/151823}, 178, 623

\bibitem[\protect\citeauthoryear{Valenzuela \& Remus}{Valenzuela \& Remus}{2024}]{valenzuelaStreamComeTrue2024}
Valenzuela L.~M.,  Remus R.-S.,  2024, \mn@doi [A\&A] {10.1051/0004-6361/202244758}, 686, A182

\bibitem[\protect\citeauthoryear{Weinberger et~al.,}{Weinberger et~al.}{2017}]{weinbergerSimulatingGalaxyFormation2017}
Weinberger R.,  et~al., 2017, \mn@doi [MNRAS] {10.1093/mnras/stw2944}, 465, 3291

\bibitem[\protect\citeauthoryear{White \& Rees}{White \& Rees}{1978}]{whiteCoreCondensationHeavy1978}
White S. D.~M.,  Rees M.~J.,  1978, \mn@doi [MNRAS] {10.1093/mnras/183.3.341}, 183, 341

\bibitem[\protect\citeauthoryear{White et~al.,}{White et~al.}{2005}]{whiteEDisCSESODistant2005}
White S. D.~M.,  et~al., 2005, \mn@doi [A\&A] {10.1051/0004-6361:20042068}, 444, 365

\bibitem[\protect\citeauthoryear{Wright, Lagos, Power, Stevens, Cortese  \& Poulton}{Wright et~al.}{2022}]{wrightOrbitalPerspectiveStarvation2022}
Wright R.~J.,  Lagos C. d.~P.,  Power C.,  Stevens A. R.~H.,  Cortese L.,   Poulton R. J.~J.,  2022, \mn@doi [MNRAS] {10.1093/mnras/stac2042}, 516, 2891

\bibitem[\protect\citeauthoryear{Yoon \& Lim}{Yoon \& Lim}{2020}]{yoonFrequencyTidalFeatures2020}
Yoon Y.,  Lim G.,  2020, \mn@doi [ApJ] {10.3847/1538-4357/abc621}, 905, 154

\bibitem[\protect\citeauthoryear{Zwicky}{Zwicky}{1956}]{zwickyMultipleGalaxies1956}
Zwicky F.,  1956, Ergebnisse der exakten Naturwissenschaften, 29, 344

\bibitem[\protect\citeauthoryear{{van~de~Sande} et~al.,}{{van~de~Sande} et~al.}{2019}]{vandesandeSAMIGalaxySurvey2019}
{van~de~Sande} J.,  et~al., 2019, \mn@doi [MNRAS] {10.1093/mnras/sty3506}, 484, 869

\bibitem[\protect\citeauthoryear{{van den Bosch}}{{van den Bosch}}{2002}]{vandenboschUniversalMassAccretion2002}
{van den Bosch} F.~C.,  2002, \mn@doi [MNRAS] {10.1046/j.1365-8711.2002.05171.x}, 331, 98

\makeatother
\end{thebibliography}

%\begin{thebibliography}{99}
%\end{thebibliography}

%%%%%%%%%%%%%%%%%%%%%%%%%%%%%%%%%%%%%%%%%%%%%%%%%%

%%%%%%%%%%%%%%%%% APPENDICES %%%%%%%%%%%%%%%%%%%%%

\appendix

\section{Completeness test}
\label{app:completeness}

\begin{figure}
    \centering
    \includegraphics[width=\linewidth]{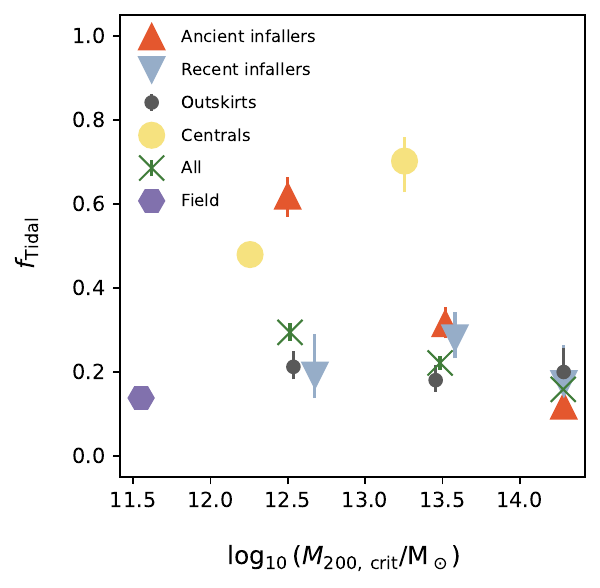}
    \caption{The tidal feature fraction as a function of the median group/cluster halo mass for satellite galaxies from the mass matched \citet{khalidCharacterizingTidalFeatures2024} sample, separated into the zones for recent infaller (blue points, downward pointing triangle), ancient infaller (red points, upward pointing triangle) and outskirt galaxies with $R_\mathrm{200,\:crit}<R_\mathrm{proj}<3\:R_\mathrm{200,\:crit}$ (black circular points), central galaxies (yellow circles), and all satellite galaxies (green crosses). The galaxies are binned by halo mass, with bin edges $\log_{10}(M_\mathrm{200,\:crit}/$M$_\odot)=[12,\:13,\:14,\:14.7]$. We plot the median halo mass and $1\sigma$ binomial confidence levels for the tidal feature fractions in each halo mass bin. The results remain qualitatively identical to Fig.~\ref{fig:tidal_frac_m200_combined}, indicating that the reduction in sample completeness does not significantly change our results.}
    \label{fig:completeness_frac_m200}
\end{figure}

To test our sample completeness, we utilise the less complete mass-matched sample from \citet{khalidCharacterizingTidalFeatures2024}. In \citet{khalidCharacterizingTidalFeatures2024}, we produced a mass-matched sample of 1300 galaxies from each simulation with identical distributions in stellar mass. By construction, this sample is a subsample of the catalogue used in this study and therefore will be significantly less complete. In the mass-matched sample, we have 543, 488 and 481 satellite galaxies in EAGLE, TNG and Magneticum, respectively. This gives us a sample $\sim71\%$ of the size of the total larger sample used in this study.

Fig.~\ref{fig:completeness_frac_m200} is similar to Fig.~\ref{fig:tidal_frac_m200_combined} in construction, but shows the less complete, mass-matched sample of satellite galaxies. However, the result remains virtually identical to Fig.~\ref{fig:tidal_frac_m200_combined}. Furthermore, the other results throughout the paper remain qualitatively similar.

\section{Richness estimates for our groups and clusters}
\label{app:richness}

\begin{figure}
    \centering
    \includegraphics[width=\linewidth]{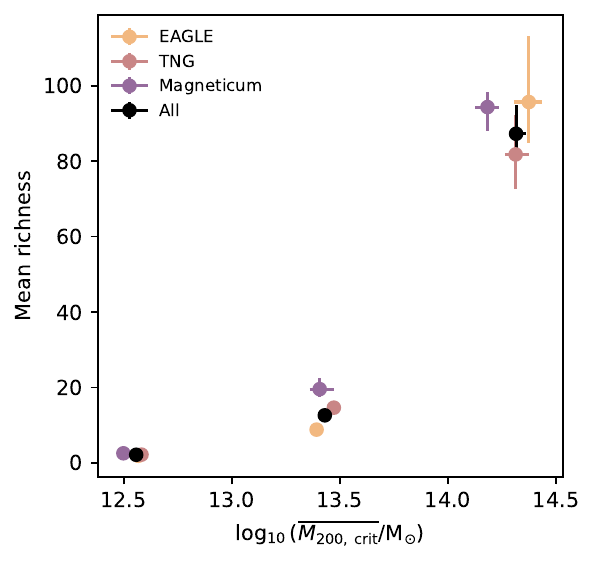}
    \caption{Mean richness as a function of mean $M_\mathrm{200,\:crit}$, for each of our halo mass bins, $\log_{10}(M_\mathrm{200,\:crit}/\mathrm{M}_\odot)=[12,\:13,\:14,\:14.7]$. We calculate the richness for all satellites with $M_\star\geq10^{9.5}$ M$_\odot$. The results for EAGLE are given in orange, TNG in red, Magneticum in purple, and we provide the means for the combined sample in black. The points and error bars show the $16^\mathrm{th}$, $50^\mathrm{th}$ and $84^\mathrm{th}$ percentiles from bootstrapping.}
    \label{fig:richness}
\end{figure}

We measure the average richness for each simulation across the $\log_{10}(M_\mathrm{200,\:crit}/\mathrm{M}_\odot)$ bins used in our work. We calculate the richness of each FOF identified group (halo) in our sample by counting the number of satellite galaxies (subhaloes) assigned to that group within 3 $R_\mathrm{200,\:crit}$ of the group/cluster centre. We show the mean richness as a function of mean group/cluster halo mass in Fig.~\ref{fig:richness}. Fig.~\ref{fig:richness} is used to estimate values for our toy model in Section \ref{subsec:implications}. 

\section{Infall times in EAGLE}
\label{app:infall_times}

We use the infall time data for EAGLE satellites from \citet{wrightOrbitalPerspectiveStarvation2022} to test if the \citet{pasqualiPhysicalPropertiesSDSS2019} zones in the projected phase-space corresponded to similar infall times for our sample. We use the data from the $z=0$ snapshot with satellite stellar masses $M_\star\geq10^{9}$ M$_\odot$, and parent halo masses $M_\mathrm{200,\:crit}\geq5.3\times10^{13}$ M$_\odot$ as this allows for a larger sample that is comparable to the sample of \citet{pasqualiPhysicalPropertiesSDSS2019}. \citet{wrightOrbitalPerspectiveStarvation2022} defines the infall time identically to \citet{pasqualiPhysicalPropertiesSDSS2019}, as the lookback time at which the satellite first came within $R_\mathrm{200,\:crit}$ of the centre of the parent halo. Our sample contains 956 satellites with measured infall times.

\begin{figure}
    \centering
    \includegraphics[width=\linewidth]{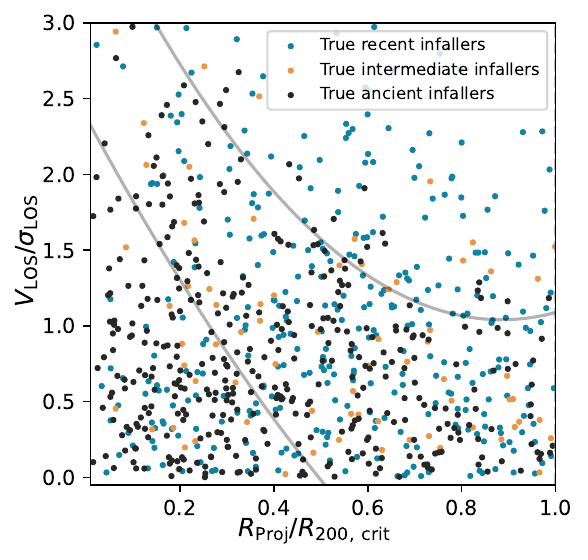}
    \caption{The projected phase-space diagram for the $z=0$ sample from \citet{wrightOrbitalPerspectiveStarvation2022}. The galaxies displayed have $M_\star\geq10^9$ M$_\odot$ and belong to parent halos with $M_\mathrm{200,\:crit}\geq10^{13}$ M$_\odot$. The points are coloured by what category of infaller they belong to based on infall time. Recent infallers are shown in blue points, first came within $R_\mathrm{200,\:crit}$ of the parent halo centre ($t_\mathrm{inf}$) $<3.36$ Gyr ago, intermediate infallers, shown with orange points, have $3.36\geq t_\mathrm{inf}/\mathrm{Gyr}<5.18$ and ancient infallers, shown with black points, have $t_\mathrm{inf}\geq5.18$ Gyr. We show using grey curves the regions used throughout the paper to divide the projected phase-space into ancient infallers (innermost region, $p\leq2$), intermediate (region between the two curves, $2<p<5$)  and recent infallers (outermost region, $p\geq5$ and $R_\mathrm{proj}\leq R_\mathrm{200,\:crit}$). The recent infaller zone has 116 galaxies, 75\% of them are recent infallers, 9\% are intermediate infallers, and 16\% are ancient infallers. The intermediate zone has 454 galaxies, 42\% are recent infallers, 10\% are intermediate infallers, and 48\% are ancient infallers. The ancient infaller zone has 226 galaxies, 27\% are recent infallers, 7\% are intermediate infallers, and 66\% are ancient infallers.}
    \label{fig:Wright_22_PPS}
\end{figure}

In Fig.~\ref{fig:Wright_22_PPS}, we plot the sample from \citet{wrightOrbitalPerspectiveStarvation2022} in the projected phase-space. There are 796 galaxies with measured infall times and $R_\mathrm{proj}\leq R_\mathrm{200,\:crit}$ to compare with the results of \citet{pasqualiPhysicalPropertiesSDSS2019}. We define true recent, ancient and intermediate infallers based on the mean infall times measured in \citet{pasqualiPhysicalPropertiesSDSS2019} for $p\leq 2$, $2<p\leq5$ and $p>5\& R_\mathrm{proj}<R_\mathrm{200,\:crit}$ zones, respectively. We see the recent infaller zone is dominated by recent infallers, and the ancient infaller zone is dominated by ancient infallers. This supports the idea that the infaller categories we adopted from \citet{pasqualiPhysicalPropertiesSDSS2019} are applicable here.

We also test the similarity between the mean infall times for the zones from \citet{pasqualiPhysicalPropertiesSDSS2019} and the mean infall times of the sample of satellite galaxies from \citet{wrightOrbitalPerspectiveStarvation2022} in these zones. We show these results in Table~\ref{tab:Wright_22_infall_times}. Given the large variance in infall times in each zone, the infall times are consistent between EAGLE and the sample from \citet{pasqualiPhysicalPropertiesSDSS2019}. This indicates that it is reasonable to use the \citet{pasqualiPhysicalPropertiesSDSS2019} measured infall times as estimates for the expected infall times for our simulations in the projected phase-space.

\begin{table}
    \centering
    \caption{The average infall times from \citet{wrightOrbitalPerspectiveStarvation2022} and standard deviations for the sample of EAGLE satellites with $M_\star\geq10^9$ M$_\odot$, belonging to parent halos with $M_\mathrm{200,\:crit}\geq10^{13}$ M$_\odot$ and the corresponding values from \citet{pasqualiPhysicalPropertiesSDSS2019} to compare to. We find reasonably good agreement in the infall times between the \citet{wrightOrbitalPerspectiveStarvation2022} and \citet{pasqualiPhysicalPropertiesSDSS2019} samples.}
    \begin{tabular}{c|c|c}
    \hline
        Zone & $\overline{t}_\mathrm{inf}$ EAGLE [Gyr] & $\overline{t}_\mathrm{inf}$ Pasquali [Gyr] \\
        \hline
        1 & $5.45\pm2.73$ & $5.42\pm2.51$\\
        2 & $5.79\pm2.61$ & $5.18\pm2.6$\\
        3 & $5.06\pm2.68$ & $4.50\pm2.57$\\
        4 & $3.89\pm2.65$ & $3.89\pm2.34$\\
        5 & $3.92\pm2.73$ & $3.36\pm2.36$\\
        6 & $3.15\pm2.70$ & $2.77\pm2.29$\\
        7 & $2.74\pm2.30$ & $2.24\pm1.97$\\
        8 & $1.39\pm1.61$ & $1.42\pm1.49$\\
        Outskirts & $3.86\pm1.71$ & \\
        \hline
    \end{tabular}
    \label{tab:Wright_22_infall_times}
\end{table}

\section{Unprojected phase-space results}
\label{app:unprojected_phase_space}

\begin{figure*}
    \centering
    \includegraphics[width=\linewidth]{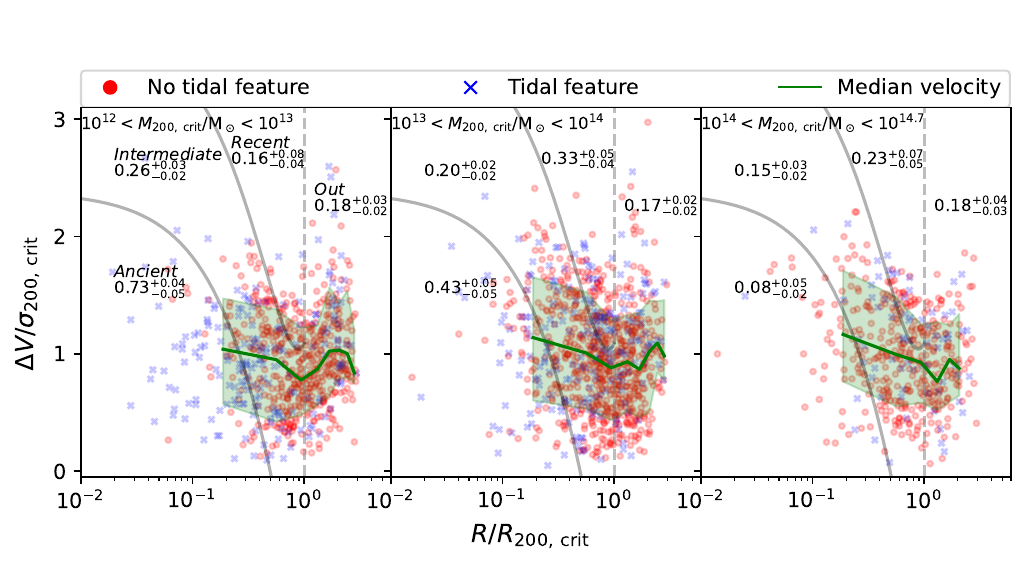}
    \caption{The unprojected velocity-radius phase-space diagram for the combined sample of satellite galaxies from EAGLE, TNG and Magneticum, across three halo mass bins $\log_{10}(M_\mathrm{200,\:crit}/\mathrm{M}_\odot)=[12,\:13,\:14,\:14.7]$. Points and lines are identical to Fig.~\ref{fig:vel_rad_tf_combined}. In this unprojected phase-space, we see a clear decreasing trend in tidal feature fractions with increasing group/cluster mass in the intermediate zone, which we did not see in the projected phase-space diagram. We also see a much clearer difference in the velocity distributions, where we now see a clear decreasing trend with increasing radius for $10^{14}<M_\mathrm{200,\;crit}/\mathrm{M}_\odot<10^{14.7}$.}
    \label{fig:vel_rad_tf_unprojected}
\end{figure*}

Fig.~\ref{fig:vel_rad_tf_unprojected} shows an identical phase-space diagram to Fig.~\ref{fig:vel_rad_tf_combined}, only for the unprojected/true separations and the norm of the 3D difference in velocities between the satellite and central galaxies. We retrieve quantitatively similar results for the recent infaller zone galaxies and the outskirt galaxies to what we saw in Fig.~\ref{fig:vel_rad_tf_combined}. Our results for ancient infaller zone satellites are qualitatively similar to Fig.~\ref{fig:vel_rad_tf_combined}, only the decrease is much sharper with a decrease from $f_\mathrm{Tidal}=0.73^{+0.04}_{-0.05}$ down to $f_\mathrm{Tidal}=0.08^{+0.05}_{-0.02}$ from the lowest to the highest group/cluster mass bins, a $\sim0.6$ decrease in tidal feature fraction as opposed to the $\sim0.47$ decrease measured in projected phase-space.

The most significant changes from Fig.~\ref{fig:vel_rad_tf_combined} are in the velocities. The decrease $\Delta V$ with increasing $R_\mathrm{proj}$ becomes clearer in the unprojected results. With a larger sample, we can reduce the impact of projection on our results. This may enable the additional results we see in the unprojected figure to emerge in the projected phase space as well.

\begin{figure}
    \centering
    \includegraphics[width=\linewidth]{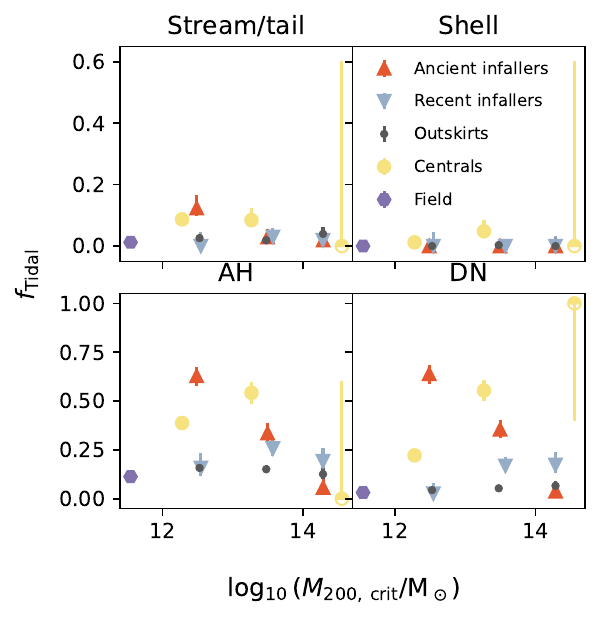}
    \caption{The specific tidal feature fraction group/cluster halo mass relationship for the combined sample of EAGLE, TNG and Magneticum, using the velocity-radius unprojected phase-space diagram to classify satellites as ancient infallers, recent infallers and outskirts galaxies. The points are identical to Fig.~\ref{fig:spec_frac_m200_combined}. We now see that the frequency of asymmetric halos and double nucleus host galaxies is clearly suppressed in ancient infallers to galaxy clusters relative to recent infallers.}
    \label{fig:spec_frac_m200_unprojected}
\end{figure}

Fig.~\ref{fig:spec_frac_m200_unprojected} shows the specific tidal feature fraction group/cluster halo mass relation where the unprojected phase-space has been used to classify satellites as ancient infallers, recent infallers and outskirt galaxies. The differences in the frequency of asymmetric halos, double-nucleus host satellites in ancient and recent infallers are more significant than when using the unprojected phase space diagram. The suppression of double nucleus and asymmetric halo hosts around ancient infaller classified satellites suggests that the tidal features from ongoing mergers are less likely within the inner regions of the cluster.

\section{Stellar mass distributions}
\label{app:v_rad_mstar}

\begin{figure*}
    \centering
    \includegraphics[width=\linewidth]{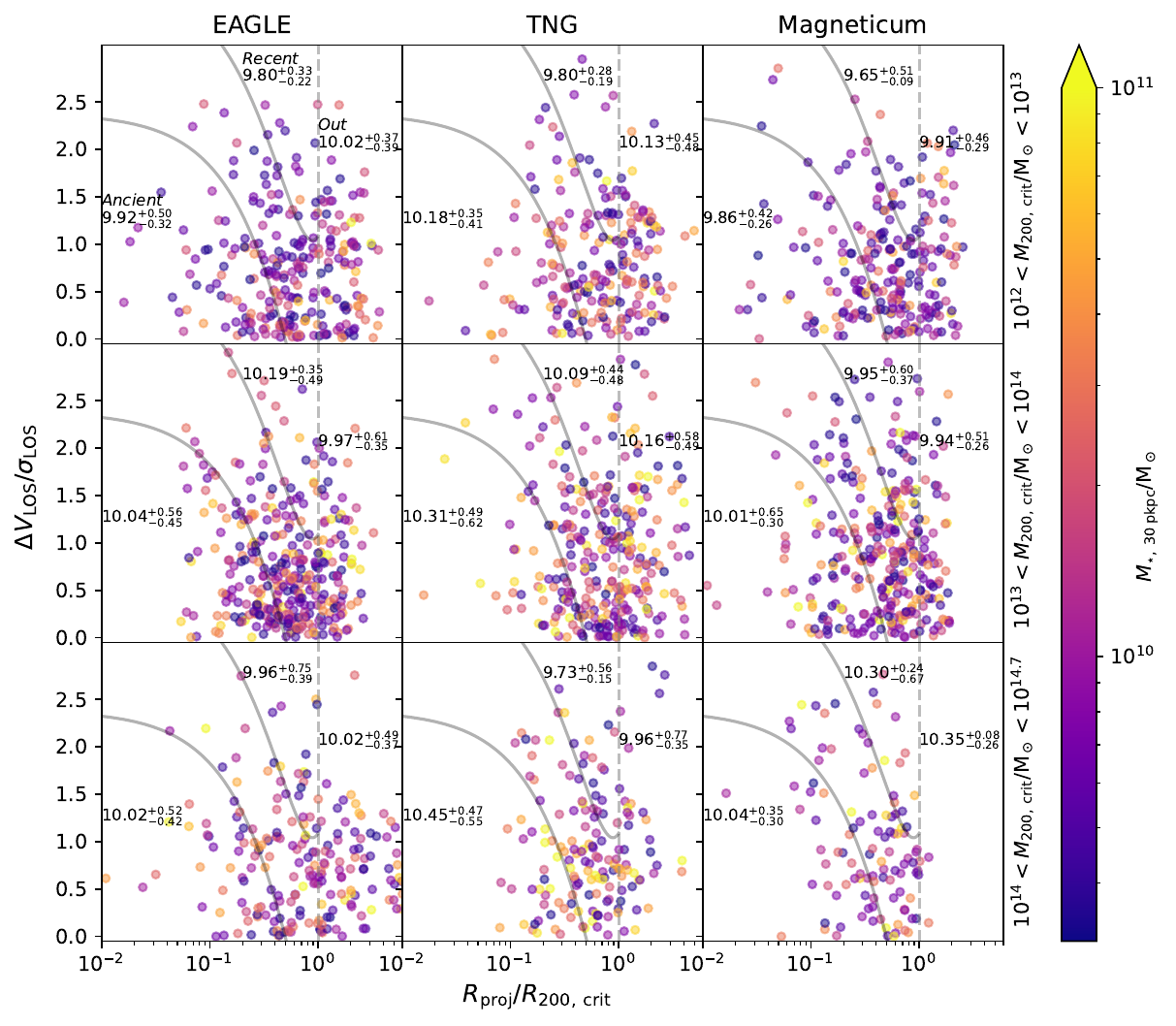}
    \caption{The projected phase-space for all satellite galaxies in a range of different halo masses for EAGLE, TNG and Magneticum. The coloured points correspond to the spherical aperture stellar masses for each of the galaxies. We show the zones from \citet{pasqualiPhysicalPropertiesSDSS2019} with $p\leq2$ that are categorised as ancient infallers, $p\geq5$ and $R_\mathrm{proj}<R_\mathrm{200,\:crit}$ are considered recent infallers, and $R_\mathrm{200,\:crit}<R_\mathrm{proj}<3\:R_\mathrm{200,\:crit}$ are the group/cluster outskirts. The median log stellar masses for each zone are provided along with the 16$^\mathrm{th}$ and 84$^\mathrm{th}$ percentiles. While the tidal feature fractions for ancient infallers decrease with increasing halo mass, the stellar masses remain consistent in each zone with increasing halo mass. We would expect a stable or increasing tidal feature fraction for ancient infallers with increasing halo mass, if galaxy stellar mass were driving the relationship between tidal feature fraction and galaxy halo mass.}
    \label{fig:vel_rad_mstar}
\end{figure*}

Fig.~\ref{fig:vel_rad_mstar} is similar to Fig.~\ref{fig:vel_rad_tf}, with the points coloured by the 30 pkpc spherical aperture stellar mass and the text showing the median log-spherical aperture stellar mass in each zone and the uncertainties giving the $16^\mathrm{th}$ and $84^\mathrm{th}$ percentiles. We can see that across the zones, there are no significant trends in stellar mass and therefore, the trends we are seeing with halo mass with respect to the tidal feature fractions for ancient and recent infallers are unlikely to be a consequence of the relationship between tidal feature fraction and stellar mass.

% If you want to present additional material which would interrupt the flow of the main paper,
% it can be placed in an Appendix which appears after the list of references.

%%%%%%%%%%%%%%%%%%%%%%%%%%%%%%%%%%%%%%%%%%%%%%%%%%

% Don't change these lines
\bsp	% typesetting comment
\label{lastpage}
\end{document}